\documentclass[prl,aps,twocolumn, floatfix, superscriptaddress,showpacs]{revtex4}
\usepackage{amsmath,bm,graphicx}
\usepackage{amssymb}
\usepackage{amsfonts}
\begin{document}

\title{Confined Brownian Ratchets}

\author{Paolo Malgaretti}
\email[Corresponding Author : ]{paolomalgaretti@ffn.ub.es }
\affiliation{Department de Fisica Fonamental, Universitat de Barcelona, Spain}
\author{Ignacio Pagonabarraga}
\affiliation{Department de Fisica Fonamental, Universitat de Barcelona, Spain}
\author{J. Miguel Rubi}
\affiliation{Department de Fisica Fonamental, Universitat de Barcelona, Spain}
\date{\today}

\begin{abstract}
 We analyze the dynamics of Brownian ratchets in a confined
environment. The motion of the particles is described by a Fick-Jakobs
kinetic equation in which the presence of boundaries is modeled by
means of an entropic potential. The cases of a flashing ratchet, a
two-state model and a ratchet under the influence of a temperature
gradient are analyzed in detail. We show the emergence of a strong
cooperativity between the inherent rectification of the ratchet
mechanism and the entropic bias of the fluctuations caused by spatial
confinement.  Net particle transport may take place in situations
where none of those mechanisms leads to rectification when acting
individually. The combined rectification mechanisms may lead to
bidirectional transport and to new routes to segregation phenomena.
Confined Brownian ratchets (CBR) could be used to control transport in
mesostructures  and to engineer new and more efficient devices for
transport at the nanoscale.
\end{abstract}


\maketitle
\section{I. Introduction}

Breaking  detailed balance due to the presence of unbalanced forces acting on a system causes rectification of thermal fluctuations and leads to new dynamical behaviors, very different from  those observed in equilibrium situations~\cite{Astumian2010,Hanggi2009}. Typical realizations of rectified motion include the transport of particles under the action of unbiased forces of optical~\cite{Mateos2011,grier}, mechanical~\cite{Balzan2011,Meer2004}, or chemical~\cite{Guerin2010,Hugel2010} origin. Hence, the implications of rectification has attracted the interest of   researchers in a variety of fields, ranging from   biology to nanoscience, due to its relevance in the transport and motion at small scales~\cite{Hanggi2009,Astumian}.  Accordingly, the behavior of such small engines, referred to as Brownian ratchets, have been deeply studied and several models that capture some of their main  features have been proposed~\cite{Astumian2010,Hanggi2009,Julicher1997,Reimann2002} . 
  
Given the small size of rectifying  elements, it is likely that  their motion takes place close to  boundaries or in confined environments. Nevertheless,  ratchet models usually assume that  rectification develops in an unbound medium. Since rectification develops as an interplay associated to how a particle experiences local forces while it explores  the space around it, confinement plays an important role because it  significantly  reduces the number of allowed  states of the rectifying elements. This restriction can be understood as an effective change of the entropy of the Brownian ratchet as it displaces along the  confined environment~\cite{Rubi2010}.
 
The relevance of entropic barriers to promote entropic transport~\cite{Reguera2006,Burada2010}  in confined environments has been  recognized in a  variety of situations that include molecular transport in zeolites~\cite{zeolites}, ionic channels~\cite{ion_channel_faraudo}, or in microfluidic devices~\cite{Bezrukov,Fujita}, where their shape explains, for example, the magnitude of the rectifying electric signal observed experimentally~\cite{hanggi_2001}. In fact, spatially varying geometric constraints provide themselves an alternative means to rectify thermal fluctuations~\cite{Rubi2010}.  Modulations in the available explored region lead to gradients
in the system effective free energy, by inducing a local bias in its
diffusion that can promote a macroscopic net velocity for
aperiodic channel profiles~\cite{Rubi2010} or due to applied alternating fields~\cite{Wambaugh}

\begin{figure}
 \includegraphics[scale = 0.28,angle=0]{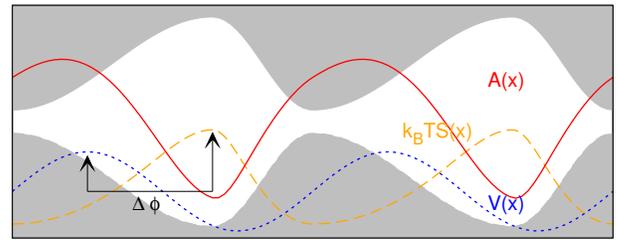}
 \caption{Brownian ratchet and entropic barriers.  A Brownian motor moving in a confined environment will be sensitive to the free energy $A(x)$ (solid) generated by the ratchet potential  $V(x)$ (dotted) and the entropic potential  (dashed), $-k_B T S(x)$, induced  by the channel shape.
\label{fig:energetic_entropic_barrier}
}
\end{figure}

We  will analyze the interplay between  rectification and confinement, and will characterize the new features associated to confined Brownian ratchets (CBR). We
show that the presence of strong cooperative rectification~\cite{Malgaretti2012}  between
both may lead to rectification even when none of
them can rectify the particle current on their own.
Such an interplay strongly affects particle motion. 
To understand the mutual influence between both rectifying sources, we
will analyze three different ratchet models, namely the flashing
ratchet, the two-level ratchet and a thermal ratchet. In the first two
cases, the equilibrium is broken by the energy injected in the system through the intrinsic ratchet mechanism as it happens in the case of molecular motors.
In the third one, the driving force is a thermal gradient that couples to the probability current, hence inducing a, local, Soret effect.

The article is distributed as follows. In Section II, we present the
main features of entropic transport and formulate the kinetic
framework, based on the  Fick-Jacobs equation, that will allow us to describe the evolution of the probability distribution of a  particle in the presence of  free energy barriers. The
ratchet models that will be analyzed in detail are described in Section III. In sections IV-VII we
discuss the different scenarios generated by different
symmetric/asymmetric ratchets and/or channel shapes, and conclude in  section VIII where 
we draw the main conclusions and outlook  of this piece of work.

\section{II. Particle dynamics in a confined medium}

A Brownian ratchet, with diffusion constant D, under the action of a potential, $V({\bf r},t)$, moving  in a confined environment characterized by a varying cross-section channel of width, $h(x,z)$, such as the one depicted in Fig.~\ref{fig:energetic_entropic_barrier},  can be characterized in terms of the probability distribution function (pdf), $P({\bf r},t)$, which obeys the Smoluchowsky equation 
\begin{equation}
 \frac{\partial}{\partial t} P({\bf r},t)=\nabla \cdot [\beta D \nabla W({\bf r})P({\bf r},t)+D \nabla P({\bf r},t)]
\end{equation}
where  $\beta^{-1}=k_BT$ is the inverse of the temperature, $T$, at which the particle diffuses, while $k_B$ stands for Boltzmann constant.  Instead of being regarded as an explicit boundary condition, the  geometrical constraint  can be included, alongside any additional potential the diffusing particle may be subject to, as an effective potential
\begin{eqnarray}
 W({\bf r}) = \begin{cases}
 V(x), & |y|\le h(x),  \&\, |z| \le L_z \\
 \infty, &  |y|> h(x)\, \mbox{or}\, |z|>L_z 
 \end{cases}
 \end{eqnarray}
where we have considered, without lack of generality, that the long axis of the channel coincides with the axis $x$, that particles cannot  penetrate the  confining channel walls, and that the channel is periodic, $W({\bf r}) = W({\bf r}+L {\bf e}_x)$, of length $L$, as shown in Fig.~\ref{fig:energetic_entropic_barrier}, and has a finite section. If the channel width varies slowly, $\partial_x h \ll 1$, one can assume that the particle equilibrates in the transverse section on time scales smaller than the ones when the particle experiences the  variations in  channel section.  It is then possible to factorize the pdf
\begin{eqnarray}
P({\bf r},t) & = & p(x,t)\frac{e^{-\beta W({\bf r})}}{e^{-\beta A(x)}} \\
e^{-\beta A(x)} & = & \int_{-L_z}^{L_z}\int_{-h(x)}^{h(x)}e^{-\beta W({\bf r})}dy dz. 
\label{normal-split}
\end{eqnarray}
By integration over the channel section one arrives at
\begin{equation}
 \frac{\partial}{\partial t} p(x,t)=\partial_x\left\{ D\left[ \beta p(x,t)\partial_x A(x)+\partial_x p(x,t)\right] \right\},
\label{FJ1}
\end{equation}
\noindent the Fick-Jacobs equation~\cite{jacobs,zwanzig,Reguera2001},  an effective one-dimensional Smoluchowsky equation that  determines the particle diffusion along the  channel.  Such motion is characterized by  the local effective free energy
\begin{equation}
 A(x)= V(x)- k_BT \ln[2 h(x)]
\label{free-en}
\end{equation}
 where $S(x)=\ln(2 h(x))$ accounts for the entropic contribution due to  confinement.
One can identify an  entropy barrier, 
\begin{equation}
 \Delta S=\ln \left(\frac{h_{max}}{h_{min}}\right)
\label{entropy-barrier}
\end{equation} 
in terms of the maximum, $h_{max}$, and minimum, $h_{min}$ channel apertures. Therefore, $\partial_x A(x)$ is the driving force that  contains entropic, $\partial_x S(x)$, and enthalpic, $\partial_x V(x)$, contributions.
The range of validity of the Fick-Jacobs equation has been analyzed~\cite{Reguera2006,Burada2007}, and it has been found that introducing the  varying diffusion coefficient~\cite{Reguera2001} 
\begin{equation}
 D(x)=\frac{D_0}{\left[1+\left(\frac{\partial  h}{\partial x}\right)^2\right]^\alpha}
\label{diff-coeff}
\end{equation}
with $\alpha=1/3,1/2$ for $3D,2D$  respectively, with the reference diffusion $D_0=k_BT/\gamma(R)$ and $\gamma(R)\propto R$, enhances the range of validity  of the factorization assumption, Eq.~(\ref{normal-split}). Although we will keep $D(x)$ for completeness, the results do not change qualitatively if a constant diffusion coefficient, $D_0$, is considered instead.

The free energy difference over a channel period   
\begin{equation}
 \Delta F=\int_0^L \partial_x A(x)dx
\end{equation}
governs the  particle current onset. Looking for the steady solution of eq.~\ref{FJ1} in a periodic system, $p_{st}(0)=p_{st}(L)$, we find that a net current, $J\neq 0$, arises  only when $\Delta F \neq 0$, which can have both an enthalpic, $\int_0^L V(x)dx\ne 0$, and entropic,  $k_BT \int_0^L \ln(h(x))dx\ne 0$, origin. Indeed the picture of  $A(x)$ as a free energy is suggestive: a net current sets only when the difference in free energy along the period is not vanishing. Clearly, at equilibrium, periodic potentials, $V(x)$, in periodic channels, $h(x)$, do not give rise to any difference in the free energy and consequently no current.  The relative performance of a  ratchet  in an uniform channel can be quantified in terms of the dimensionless parameter
\begin{equation}
 \mu_0=\frac{L v_0}{\tilde{\mu} \Delta F_0}
\label{mobility0}
\end{equation}
defined as the ratio between the Brownian ratchet average speed, $\bar{v}=\frac{1}{L}\int_0^LJ(x)dx$ with $J(x)=D\left[ \beta p(x,t)\partial_x A(x)+\partial_x p(x,t)\right]$  derived from  eq.~\ref{FJ1}, and the average speed of a particle with mobility $\tilde{\mu}\equiv \beta D_0$ under the action of a uniform effective force,  $f_0\equiv \Delta F_0/L$. In the absence of intrinsic ratchet rectification, $\Delta F_0=0$ and $\mu_0$ remains 1. If the ratchet leads to an intrinsic rectification, the interplay   between  ratcheting and  confinement can be alternatively quantified interns of the dimensionless parameter
\begin{equation}
 \mu=\frac{L \bar{v}}{\tilde{\mu} \Delta F}
\label{mobility}
\end{equation}
that accounts for the overall free energy drop $\Delta F$.
 For a uniform channel, when rectification is purely enthalpic, $\mu/\mu_0=1$. Therefore, deviations of $\mu/\mu_0$ from 1  constitute a convenient means to address the role of entropic constraints to particle rectification; $\mu/\mu_0 > 1$ indicates that the geometrical constraints  cooperate with the force associated to the Brownian ratchet to induce an efficient cooperative rectification, larger than the one obtained in an unstructured environment, while the opposite holds for $\mu/\mu_0 < 1$. The absolute value of $\mu$ gives additional information. On one hand, when $\mu>1$ the performance of the cooperative rectification beats the one obtained under a constant force $f\equiv \Delta F/L$, on the other hand $\mu$ also  easily identifies  the non-linear rectifying regime, when $\partial_{f}\mu\ne0$. 

To analyze the interplay between the Brownian ratchet and confinement, we will consider that all Brownian ratchets are subject to the same underlying, driving periodic potential,
\begin{equation}
V_0({\bf r}) = V_0 \left[\sin  \frac{2 \pi}{L} x +\lambda \sin \frac{4 \pi}{L} x \right]
\end{equation}
This is a simple potential, explored in detail previously in which rectification is controlled  by a single  parameter, $\lambda$~\cite{reimann_appl_2002}. We will consider a channel width that varies with  the same periodicity as $V_0$ and  with a similar  functional dependence
\begin{equation}
 h(x)=h_0-R+h_1 \sin \left[\frac{2\pi}{L}(x+\phi_0)\right]+h_2\sin\left[\frac{4\pi}{L}(x+\phi_0)\right]
\end{equation}
where $h_0$ is the average channel section and $h_1$ and $h_2$ determine its modulation. In particular $h_2$ is responsible for the symmetry of the channel along its transverse axis. When $h_2 \neq 0$ the left-right symmetry along the channel longitudinal axis is broken while, for $h_2=0$, the left-right symmetry of channel along its longitudinal axis is restored. $h_{max}$ and $h_{min}$ depend both on $h_1$ and $h_2$ and the dephasing, $\phi_0$. The latter will be useful to  displace the geometrical and potential modulations, as discussed in the next Sections. The particle radius, $R$, affects the available  transverse section and will hence contribute to the entropic barrier, Eq.(~\ref{entropy-barrier}).

\section{III. Ratchet models}
In order to analyze the impact of confinement in the rectification of a Brownian particle, we will consider three different types of complementary, well-established  ratchet models that have the same periodicity that the  geometric confinement. 

\subsection{Flashing ratchet}
A colloidal particle subjected to a periodic external potential 
\begin{equation}
 V(x)=\mathcal{V}_1V_0(x)
\end{equation}
 behaves as a flashing ratchet when the random force breaks detailed balance~\cite{grier}. This can be simply achieved for a Gaussian  white noise with a second moment amplitude $g(x)=\sqrt{D(x)+Q\left(\partial_{x}V_0(x)\right)^{2}}$~\cite{Reimann2002}, where $Q$ controls  Brownian rectification. The Fick-Jacobs equation for such a  flashing ratchet in a varying-section channel reads
\begin{equation}
 \frac{\partial}{\partial t} p(x)  = \frac{\partial}{\partial x}\left\{g(x)\frac{\partial [p(x)g(x)]}{\partial x}+ D(x) p(x)\frac{\partial\beta A(x)}{\partial x}\right\},
\label{FJ-Reimann}
\end{equation}
which reduces to equilibrium diffusion for $Q=0$. For $Q>0$ detailed balance is broken and net fluxes  arise  when 
\begin{equation}
 \beta\Delta F=\int_{0}^{L}\left[\frac{D(x)\partial_x A(x)}{g(x)^2}+\frac{\partial}{\partial_x}\ln g(x)\right]dx\ne 0.
\label{free-energy-Reimann}
\end{equation}
 Since  $\int_0^L\partial_x\ln g(x)dx=0$ for a periodic channel,  particle currents emerge from the interplay between both the entropic and enthalpic forces, encoded in $A(x)$ and the position-dependent noise, $g(x)$. Three dimensionless parameters governs the Brownian ratchet performance: $\beta \mathcal{V}_1$ and $\Delta S$ quantify  the relevance of the enthalpic and entropic contributions, respectively, while $Q/(L^2D_0(R))$ determines  rectification. 

\subsection{Two state molecular motor}
The two-state ratchet model constitutes a standard, simple framework to describe  molecular motor motion. A  Brownian particle jumps between two states,  $i=1,2$, (strongly and  weakly bound) that determine under which potential, $V_{i=1,2}$, it displaces~\cite{Julicher1997}.  A choice of the jumping rates $\omega_{12,21}$ that break detailed balance, jointly with an asymmetric potential  of the bound sate, $V_{1}(x)$, determines the average molecular motor velocity $v_0\neq0$.  The conformational changes of the molecular motors  introduce an additional  scale that will compete with  rectification and geometrical confinement.  Infinitely-processive molecular motors remain always attached to the filament along which they displace and are affected by the geometrical restrictions only while displacing along the filament; accordingly, we choose  channel-independent binding rates $\omega_{21,p}(x) = k_{21}$. On the contrary, highly non-processive molecular motors detach frequently  from the  biofilament  and 
diffuse away; an effect we account for considering a   channel-driven binding rate, $\omega_{21,np}(x)=k_{21}/h(x)$. 
Motors jump to the weakly bound state only in a region of width $\delta$ around the  minima of  $V_1(x)$,  with rate $\omega_{12}  =   k_{12}$. Accordingly, the motor densities in the strong(weak) states, $p_{1(2)}$ along the channel follow~\cite{Julicher1997}
\begin{eqnarray}
\partial_{t}p_{1}(x)+\partial_{x}J_{1} & = & -\omega_{12}(x)p_{1}(x)+\omega_{21}(x)p_{2}(x) \nonumber\\
\partial_{t}p_{2}(x)+\partial_{x}J_{2} & = & \omega_{12}(x)p_{1}(x)-\omega_{21}(x)p_{2}(x) 
\label{FJ-two-states}
\end{eqnarray}
where $J_{1,2}(x)  =  -D(x)\big[\partial_{x}p_{1,2}(x)+p_{1,2}(x)\partial_{x}\beta A_{1,2}(x)\big]$ stands for the current densities in each of the two states in which motor displaces. Depending on the motor internal state, two  free energies,  $A_{1,2}(x)=V_{1,2}(x)-k_BT S(x)$, account for the interplay between the biofilament interaction and the channel constraints.  Since molecular motors can jump between two internal states, the corresponding expression for the overall free energy drop, $\Delta F$,  must be generalized to account  for these internal changes. Accordingly, the overall free energy drop must be identified from a two dimensional  particle flux,  a treatment that we leave for a future study. This system is also characterized by three dimensionless parameters:  $\beta \mathcal{V}_1$ and $\Delta S$ control the amplitude of the enthalpic  and entropic contribution, respectively, while $\omega_1/\omega_2$  quantifies the departure from detailed balance. 

\subsection{Thermal ratchet} As a last example, we will consider  a Brownian particle moving in a varying-section channel under the influence of a local temperature gradient. In the absence of entropic barriers, the transport induced by the imposed temperature gradient  has been analyzed previously~\cite{vanKampen,Buttiker}. By assuming local equilibrium along the radial directions the probability distribution function obeys~\cite{zwanzig}: 
\begin{equation}
 P({\bf r},t)=p(x,t)\frac{e^\frac{-W({\bf r})}{K_BT(x)}}{e^\frac{-A(x)}{k_BT(x)}}
\label{temp-pdf-split}
\end{equation}
and the corresponding  Fick-Jacobs equation reads
\begin{eqnarray}
 &\frac{\partial }{\partial t}p(x,t)=\frac{\partial}{\partial x}\bigg\{\mu(x)p(x,t)\bigg(T(x)\frac{\partial}{\partial x}\frac{A(x)}{T(x)}+\bigg.\bigg.\nonumber\\
&\bigg.+V(x)\frac{\partial}{\partial x} \ln T(x)\bigg)\bigg.+\mu(x)\frac{\partial}{\partial x} [ T(x)p(x,t) ]\bigg\},
\end{eqnarray}
where  we keep the phenomenological  dependence of the diffusion on the channel width through  the local mobility $ \mu(x)=\beta D(x)$.
The overall free energy drop can be expressed as
\begin{equation}
 \Delta F=\int_0^L \left[\frac{\partial}{\partial x}  \frac{A(x)}{T(x)} +\left(\frac{V(x)}{T(x)}+1\right) \frac{\partial}{\partial x} \ln T(x) \right]dx.
\label{free-energy-temp}
\end{equation}
which differs qualitatively from the one obtained for the flashing ratchet,  eq.~\ref{free-energy-Reimann}. For a periodic thermal ratchet under a periodic temperature gradient,  the entropic contribution, $ \int_0^L \frac{\partial }{\partial x} \frac{A(x)}{T(x)} dx$, vanishes and does not contribute to $\Delta F$. Therefore, rectification in a periodic thermal ratchet, quantified by $\Delta F= \int_0^L\frac{V(x)}{T(x)}\partial_x \ln T(x)dx$, can only develop  from an interplay between the enthalpic and temperature  variations~\cite{Buttiker, Mazur}. 
Although both the flashing and the thermal ratchet are characterized by a multiplicative noise, their different physical origin is at the basis of this different response. For the thermal ratchet the  spatial inhomogeneity affects  both the amplitude of the fluctuations and the local  equilibrium distribution while for the flashing ratchet the noise amplitude is regarded as an effective coarse-graining of a molecular mechanism that is decoupled from the underlying equilibrium properties of the Brownian particle. In fact, if we would (inconsistently) neglect the  spatial dependence of the temperature in the equilibrium distribution of the thermal ratchet that appears in Eq.~(\ref{temp-pdf-split}), we would  derive an $ \Delta F_T$, 
\begin{equation}
 \Delta F_T=\int_0^L\frac{\partial_xA(x)}{T(x)}+\partial_x\ln T(x)dx
\end{equation}
qualitatively analogous to the one obtained for the flashing ratchet, Eq.~\ref{free-energy-Reimann}.  A similar result, and hence the possibility of  constrained-controlled rectification, is obtained if one assumes that the temperature becomes anisotropic. This corresponds to situations where the equilibration transverse to the  channel is determined by a temperature that differs from the one characterizing the particle  diffusion along the channel. Such situations can develop if there is an intrinsic mechanism for energy dissipation, as has been reported. e.g. in vibrated granular gases~\cite{reimann_epl}.

\section{IV. Fully symmetric case}
We will first consider the case of symmetric ratchet and entropic potentials,
implemented for $h_2=\lambda=0$. Under these conditions, current
rectification is not possible when entropic and enthalpic forces act
 separately. We will show in this Section that rectification may arise
due to the interplay between both drivings, when they are phase shifted
an amount $\phi_0$.

\begin{figure}
 \includegraphics[scale=0.23]{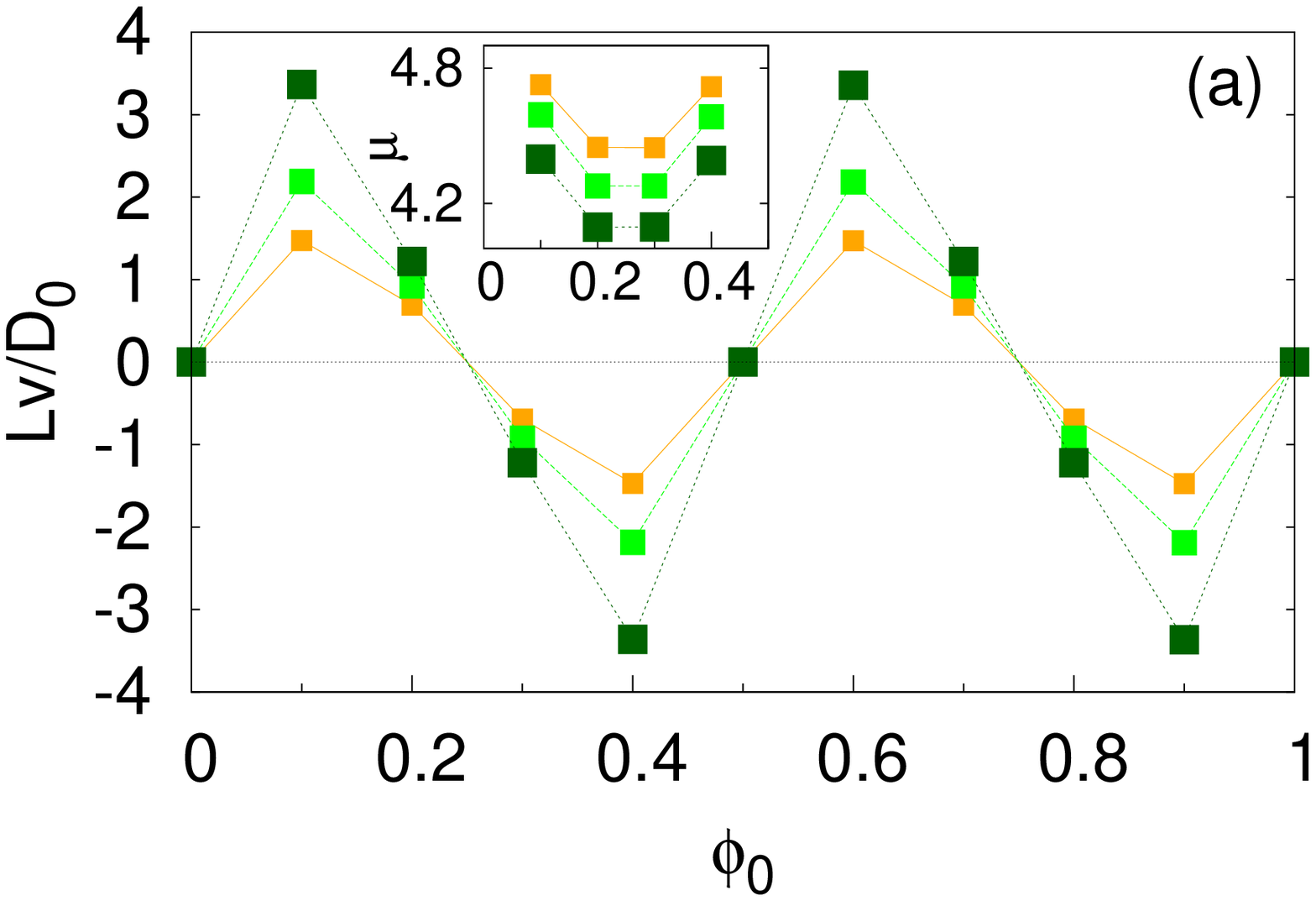}\includegraphics[scale=0.23]{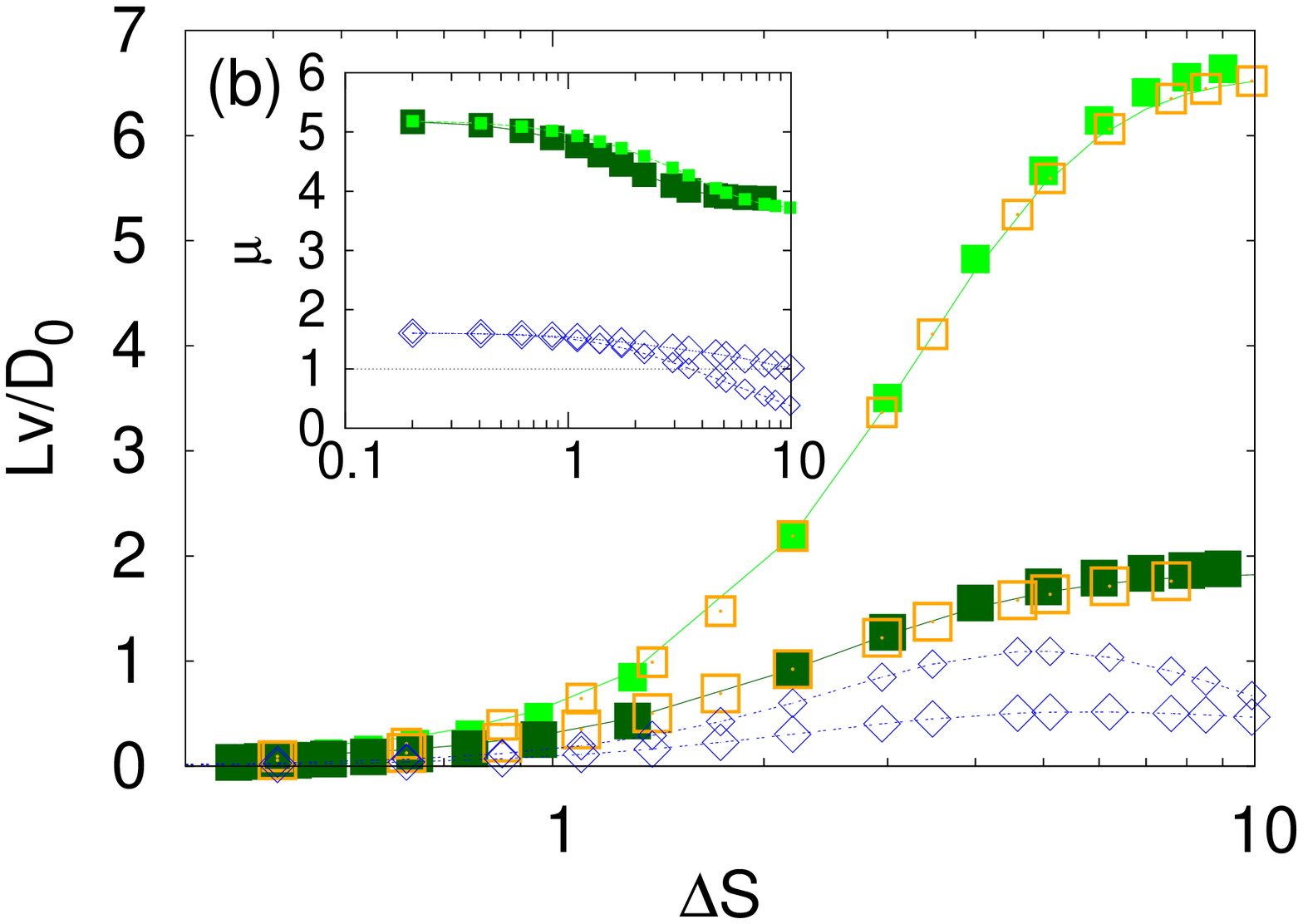}
\includegraphics[scale=0.23]{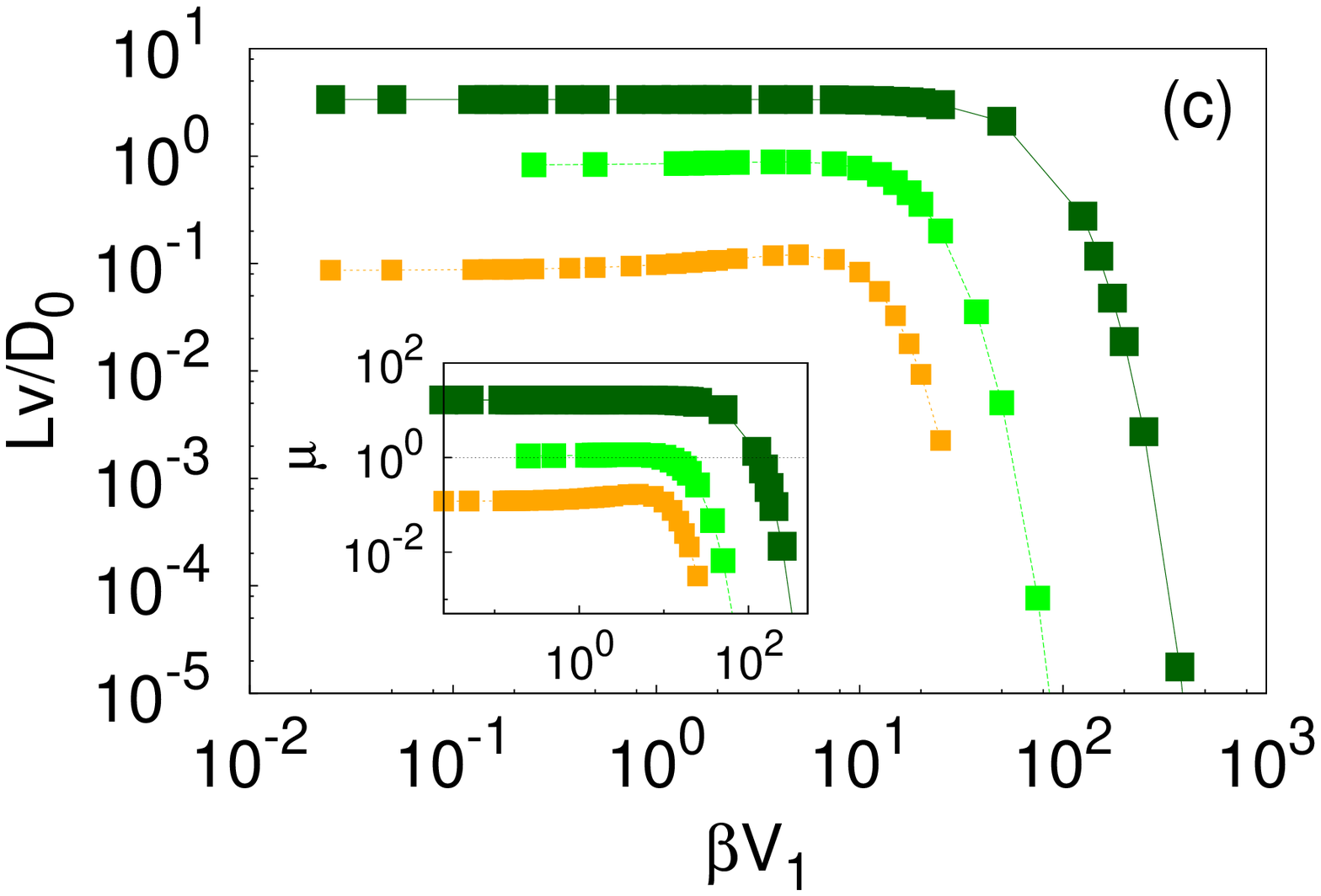}\includegraphics[scale=0.23]{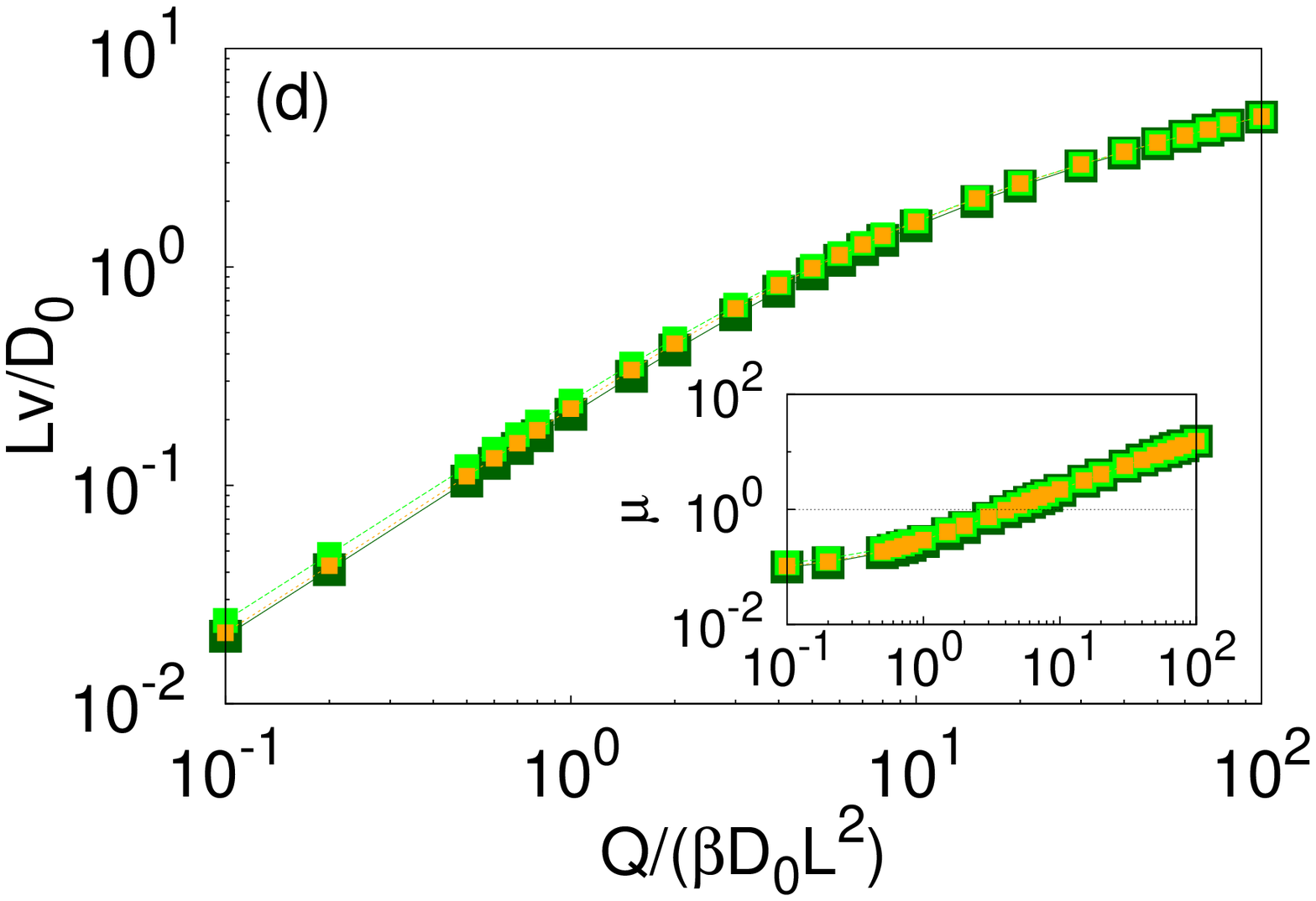}
\caption{Rectification of a Brownian motor moving due to a symmetric flashing ratchet in a symmetric channel. (a): particle velocity, in units of $D_{0}/L$, being $D_0=D_0(R=1)$, as a function of the phase shift $\phi_0$  for different values of the parameter $\Delta S=1.73,2.19,2.94$ (the larger the symbol   size, the larger $\Delta S$), being  $\mathcal{V}_1=0.2$ and $Q=2$. Inset: $\mu$ as a function of $\phi_0$ for the same parameters. (b): particle velocity as a function of $\Delta S$ upon variation of particle radius $R$ (solid lines, with $h_0=1.25,h_1=0.2$), $h_0$ (solid points, with $R=1,h_1=0.2$) or $h_1$ (open points, with $R=1,h_0=1.25$) for $\phi_0=0.1,0.2$ and $\mathcal{V}_1=0.2,Q=2$ (the larger the symbol the larger  $\phi_0$). As a comparison the case with $\mathcal{V}_1=0.2,Q=0.02$ and $\phi_0=0.1,0.2$,  (the larger the symbol the larger  $\phi_0$)., is shown (blue diamonds). Inset: $\mu$ as a function of $\Delta S$ for the same parameters. (c): particle velocity as a function of the ratchet 
potential amplitude $\mathcal{V}_1$ for $Q=0.02,0.2,2$  (the larger the symbol the larger  $Q$), while $\Delta S=2.94,\phi_0=0.1$. Inset: $\mu$ as a function of $\phi_0$ for the same parameters. (d): particle velocity as a function of $Q$ for $\mathcal{V}_1=0.02,0.2,2$ and $\Delta S=2.94,\phi_0=0.1$,  (the larger the symbol the larger  $\mathcal{V}_1$).}
\label{fig-reimann-simm-simm}
\end{figure}

\emph{Flashing Ratchet}. Fig.~\ref{fig-reimann-simm-simm}.a shows the particle current obtained by solving eq.~\ref{FJ-Reimann} under the steady-state condition $\dot{p}(x)=0$. As shown in fig.~\ref{fig-reimann-simm-simm}.a, a net particle current develops when the phase shift, $\phi_0$, is not a multiple of $\pi$. Such a particle flux is the result of the interplay between confinement and the potential leading to particle rectification. In fact, eq.~\ref{free-energy-Reimann} together with eq.~\ref{free-en} clearly show that, if the channel and the ratchet are not in phase, $\phi_0\ne0$,  the overall  free energy drop, eq.~\ref{free-energy-Reimann}, is finite and a net current develops.
The Fick-Jacobs equation identifies $\phi_0$ and $\Delta S$ as the relevant parameters that control rectification.   $\phi_0$ is  responsible for the spatial symmetry breaking~\cite{footnote1} for any finite  channel modulation, $\Delta S$. A straight channel, $\Delta S=0$, will not induce rectification because the  underlying potential is symmetric. As shown in  Fig.\ref{fig-reimann-simm-simm}.b, $\Delta S$,  quantifies the changes in the system geometrical properties  by tuning the particle radius, $R$, or the channel corrugation, $h_1$.
 
Particle current varies smoothly with the other dimensionless parameters, $\mathcal{V}_1$ and $Q$. For example, the value of $\Delta S$ providing the maximum current is only weakly affected by a drop of $Q$ of two orders of magnitude, as shown in Fig.~\ref{fig-reimann-simm-simm}.b. For small values of $Q$,  the particle velocity does not increase monotonously with  $\mathcal{V}_1$. While Fig.~\ref{fig-reimann-simm-simm}.c indicates that there exists a finite value of  $\mathcal{V}_1$ at which particle flux is optimal,  Fig.~\ref{fig-reimann-simm-simm}.d shows that the particle current increases monotonously with $Q$, leading to a monotonous increase of the particle current. 
Since both the channel corrugation and the ratchet potential are symmetric,  Fig.~\ref{fig-reimann-simm-simm}.a is symmetric under inversion of the velocity and dephasing angle. Therefore, a uniform distribution of $\phi_0$ will not induce any net current, but any asymmetric distribution will. 

The insets of Fig.~\ref{fig-reimann-simm-simm} display the changes of the dimensionless velocity,  $\mu$ (Eq.~(\ref{FJ-Reimann})), as a function of the relevant dimensionless parameters. They show that there are  regimes where confinement and the ratchet potential cooperate to induce an efficient particle rectification, $\mu > 1$, and regimes where they  compete with each other, partially hindering rectification, $\mu <1$. Both regimes depend weakly on  confinement, as shown in the top panels of fig.~\ref{fig-reimann-simm-simm}, while $\mu$ is significantly  affected by the magnitude both of the ratcheting potential,  $\mathcal{V}_1$, and the noise amplitude, $Q$. In particular, $\mu<1$ is typical for small values of $\mathcal{V}_1$ and $Q$ while, upon increasing both $\mathcal{V}_1$, and $Q$, the $\mu>1$ regime arises. As the magnitude of   $\mathcal{V}_1$ increases, $\mu$ decreases drastically and   the net current eventually vanishes. 

\begin{figure}
 \includegraphics[scale=0.23]{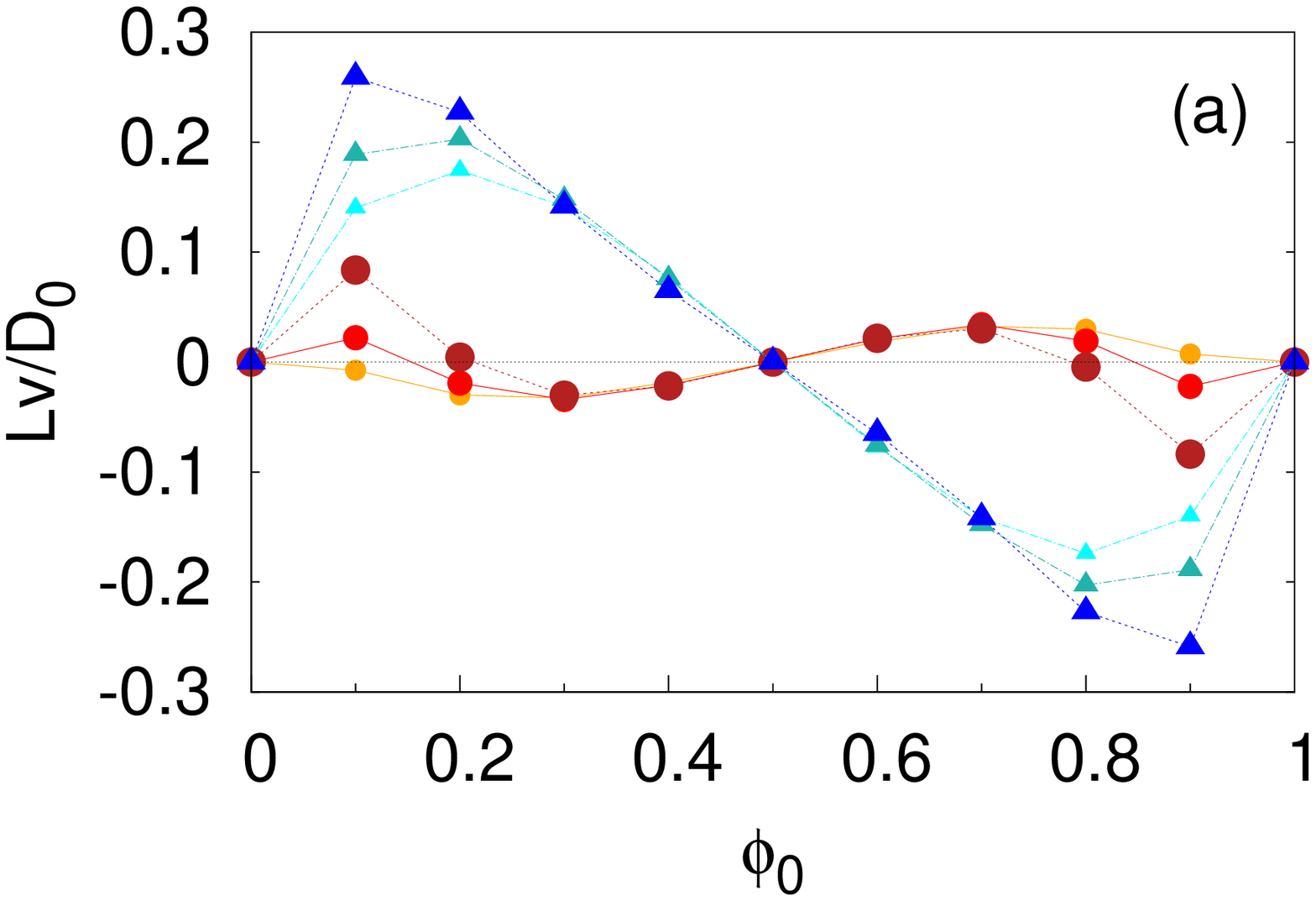}\includegraphics[scale=0.23]{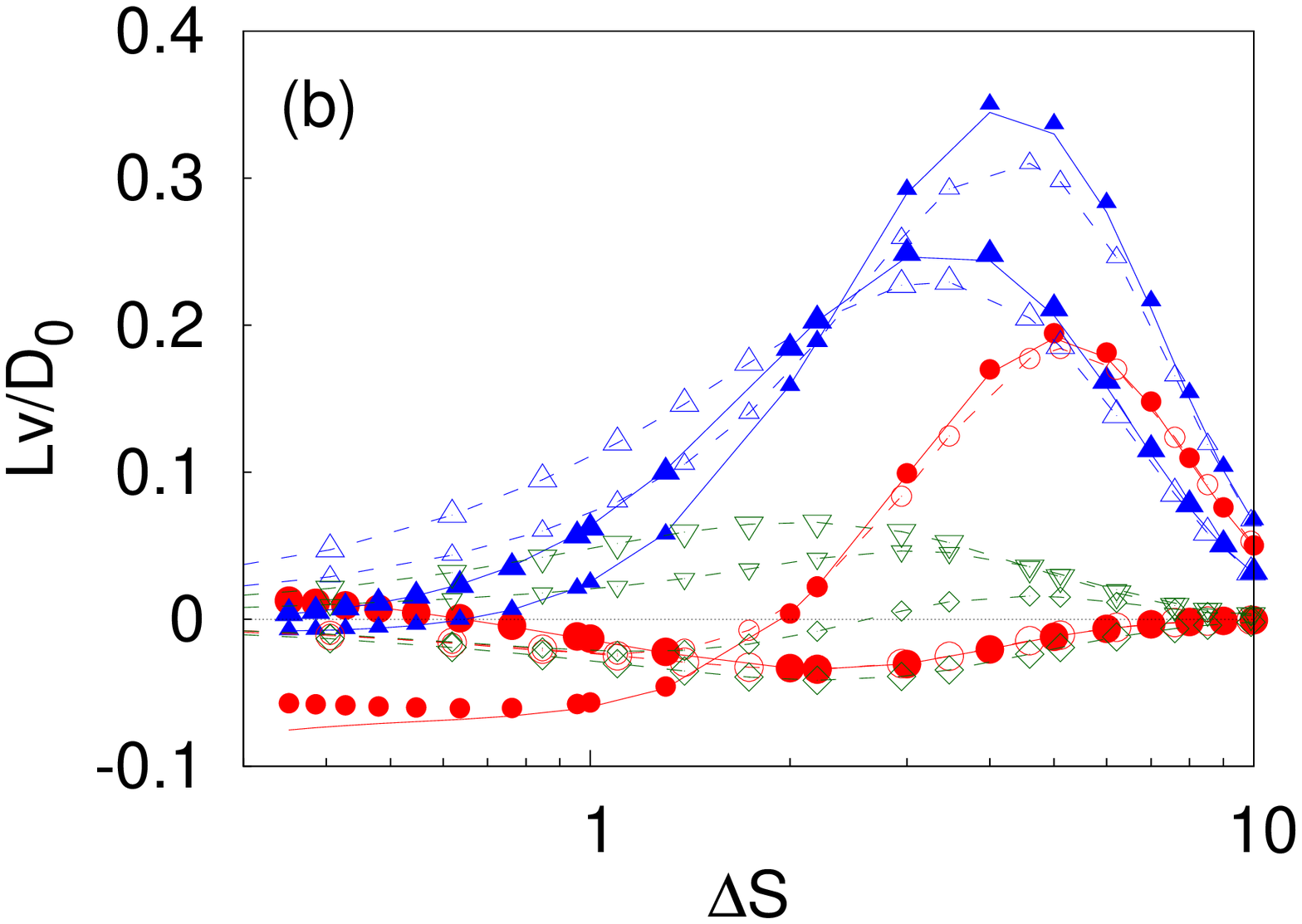}
\includegraphics[scale=0.23]{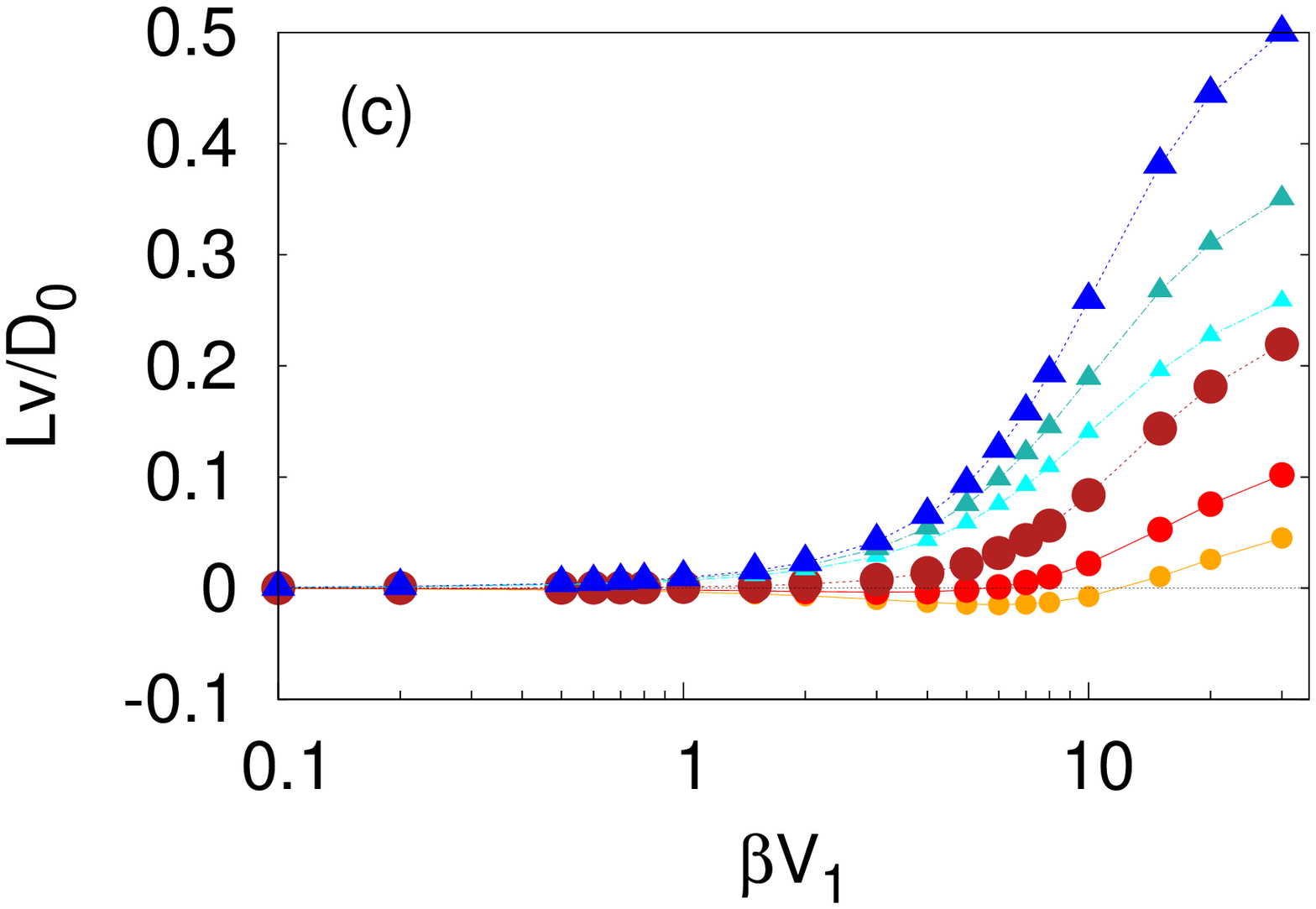}\includegraphics[scale=0.23]{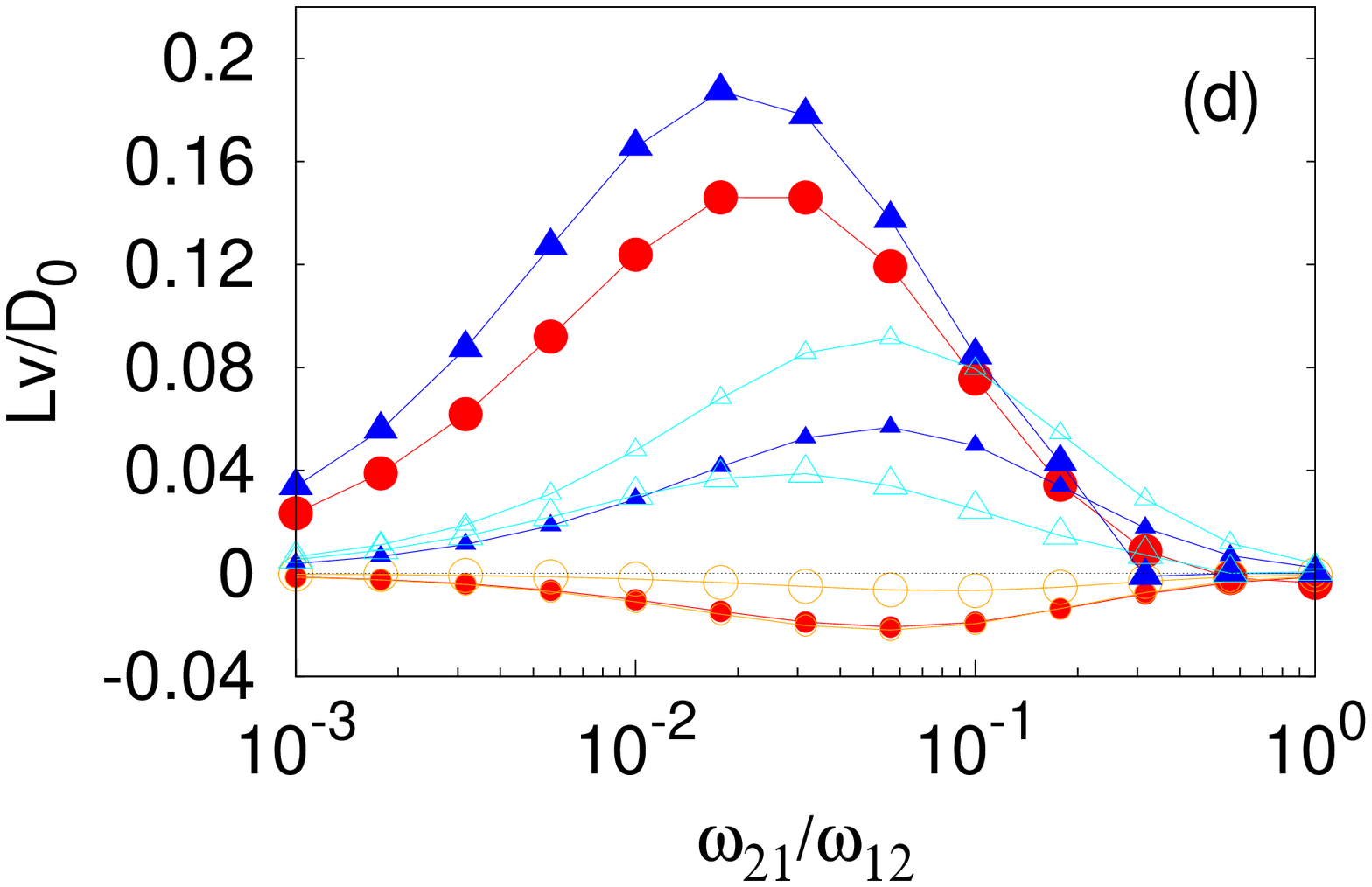}
\caption{Rectification of a processive (circles), non-processive (triangles) Brownian particle moving due to the two state model in a symmetric channel. (a): particle velocity, in units of $D_{0}/L$, with $D_0=D_0(R=1)$, as a function of the phase shift $\phi_0$  for different values of the parameter $\Delta S=1.73,2.19,2.94$ (the larger the symbol   size, the larger $\Delta S$), for  $\mathcal{V}_1=0.2$ and $\omega_{2,1}/\omega_{1,2}=0.01$. (b): Processive (circles), non-processive (triangles) Brownian motor velocity, in units of $D_{0}/L$, as a function of $\Delta S$ upon variation of particle radius $R$ (solid lines, for $h_0=1.25,h_1=0.2$), $h_0$ (solid points, for $R=1,h_1=0.2$) or $h_1$ (open points, for $R=1,h_0=1.25$) for $\phi_0=0.1,0.2$(larger symbols correspond to larger  $\phi_0$). As a comparison, the case for $\phi_0=0.1,0.2$ and $\mathcal{V}_1=0.2$ is shown (green diamonds) (the larger the symbol the larger  $\phi_0$)  with  $\omega_{2,1}/\omega_{1,2}=0.01$. (The curves for $\mathcal{V}_1=1$ 
have been magnified by a factor of $5$ for the sake of clarity.) (c): Processive (circles), non-processive (triangles) Brownian motor velocity as a function of the ratchet potential amplitude $\mathcal{V}_1$ for $\Delta S=1.73,2.19,2.94$ (the larger the symbol the larger  $\Delta S$)   with  $\omega_{2,1}/\omega_{1,2}=0.01$. (d): Processive (circles), non-processive (triangles) Brownian motor velocity, in units of $D_{0}/L$, as a function of $\omega_{1,2}/\omega_{2,1}$ for $\phi_0=0.1\,(0.3)$, open (solid) points and $\Delta S=0.4,7.6$  (the larger the symbol the larger  $\Delta S$), with $\mathcal{V}_1=2$.}
\label{fig-twostate-simm-simm}
\end{figure}
\emph{Two state model} 
Fig.~\ref{fig-twostate-simm-simm}.a  shows that a  net particle current develops when  the ratchet potential and the channel corrugation are out of registry. The symmetry of the channel and  rectifying potentials  imply that the  velocity profile is invariant if both axis of the figure are inverted; hence a uniform distribution of $\phi_0$ will not induce a net current, but any asymmetric distribution will. The internal reorganization of the molecular motor as it moves along the channel allows for qualitatively new scenarios with respect to the  rectification features observed for the flashing ratchet. For example, Figs.~\ref{fig-twostate-simm-simm}.b and \ref{fig-twostate-simm-simm}.c show that the particle flux can reverse its direction as $\Delta S$  and $\mathcal{V}_1$ increase, respectively, although flux reversal is more sensitive to channel corrugation.   This flux reversal can be exploited to  induce particle separation according to their  size (due to the implicit dependence of   $\Delta S$ on 
particle radius, $R$),  or the differential particle response to $\mathcal{V}_1$~\cite{footnote2}. 

Although $\Delta S$  captures the essential features of net molecular motor motion, Fig.~\ref{fig-twostate-simm-simm}.b shows separate sensitivity to the rest of the geometrical channel parameters,   $h_0$, $h_1$, as well as motor size  $R$. Such sensitivity is remarkable at smaller entropic barrier magnitudes  where the confining-tuned diffusion coefficient, eq.(~\ref{diff-coeff}), plays a relevant role. If  $h_1\rightarrow 0$ then both the entropic barrier and  the modulation of the diffusion coefficient  vanish.  On the contrary if  $R$ or $h_0$ decrease at fixed $h_1$, then    $\Delta S\rightarrow0$ but the   modulation of the diffusion coefficient persists, allowing for rectification.  As  $\Delta S$ increases,  the sensitivity to the separate variation of  $h_0$, $h_1$ and $R$ for the case of non-processive motors intensifies. These deviation from the geometrical dependence only through $\Delta S$ arise because the binding rate, $\omega_{2,1}$, depends on the probability that the motor is close to the 
filament, which depends indirectly on the  channel section.  Hence, different channel amplitudes $h_0,h_1,R$, even if leading to the same $\Delta S$, give rise to different binding rates that modulate  the molecular motor velocity. This sensitivity is not present for processive motors, as observed  in Fig.~\ref{fig-twostate-simm-simm}.c.

Molecular motors show a maximum current for an optimal $\Delta S$ that depends weakly on $\mathcal{V}_1$, as displayed in Fig.~\ref{fig-twostate-simm-simm}.b, while Fig.~\ref{fig-twostate-simm-simm}.d shows that  an optimum velocity, sensitive both to $\phi_0$ and  $\Delta S$, can also be achieved on increasing the ratio of binding and unbinding rates,  $\omega_{2,1}/\omega_{1,2}$ . 

\section{V. Symmetric potential and asymmetric channel}
The channel asymmetry wit respect to its transverse axis, $h_2\ne 0$, breaks the left-right spatial symmetry along the channel longitudinal axis. Such an asymmetry leads to  substantial changes in rectification with respect to the symmetric channel described earlier because now there exists a geometrically-induced preferential direction for particle rectification. As in the previous case, rectification here is induced by the asymmetric confinement.

\begin{figure}
 \includegraphics[scale=0.23]{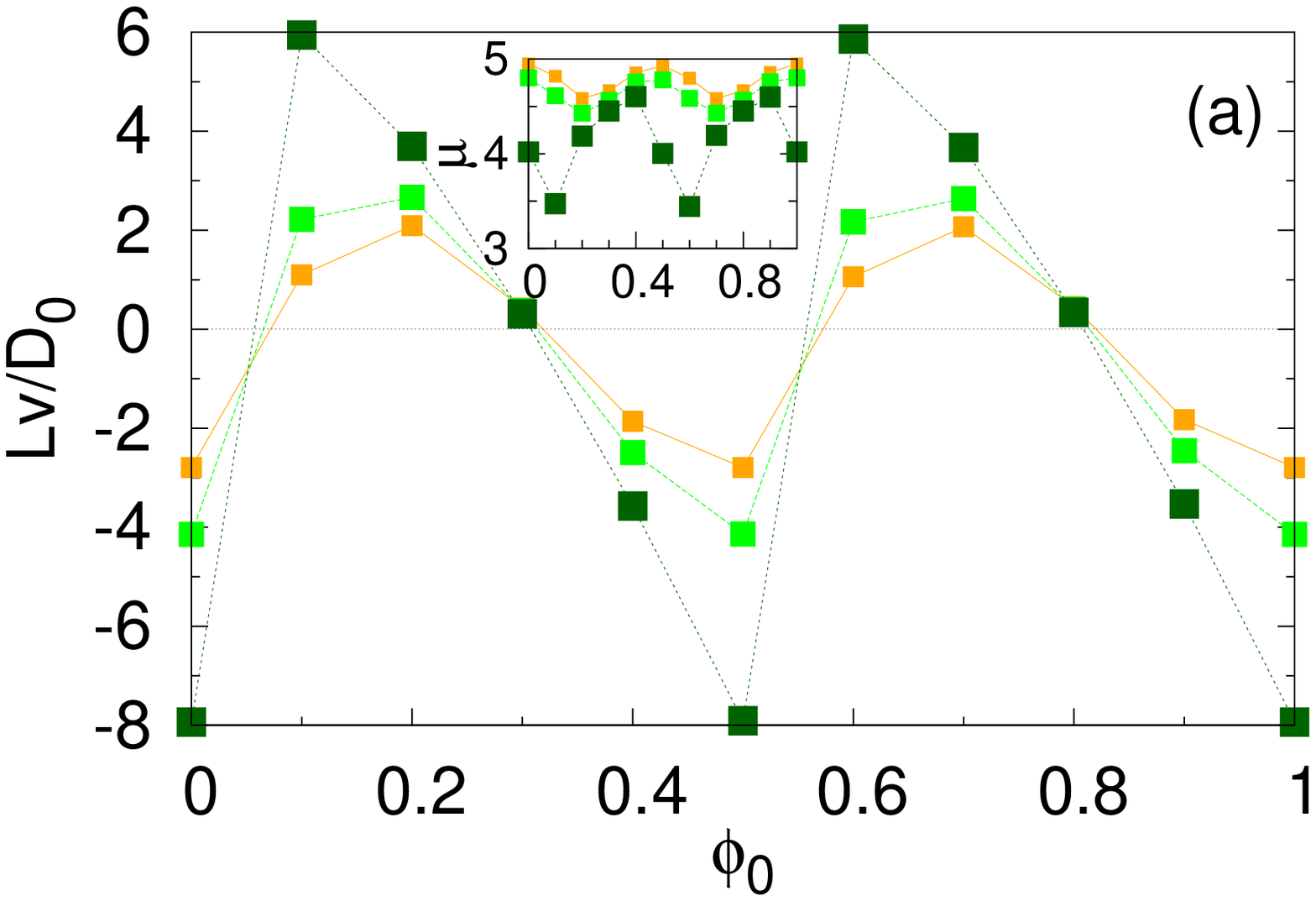}\includegraphics[scale=0.23]{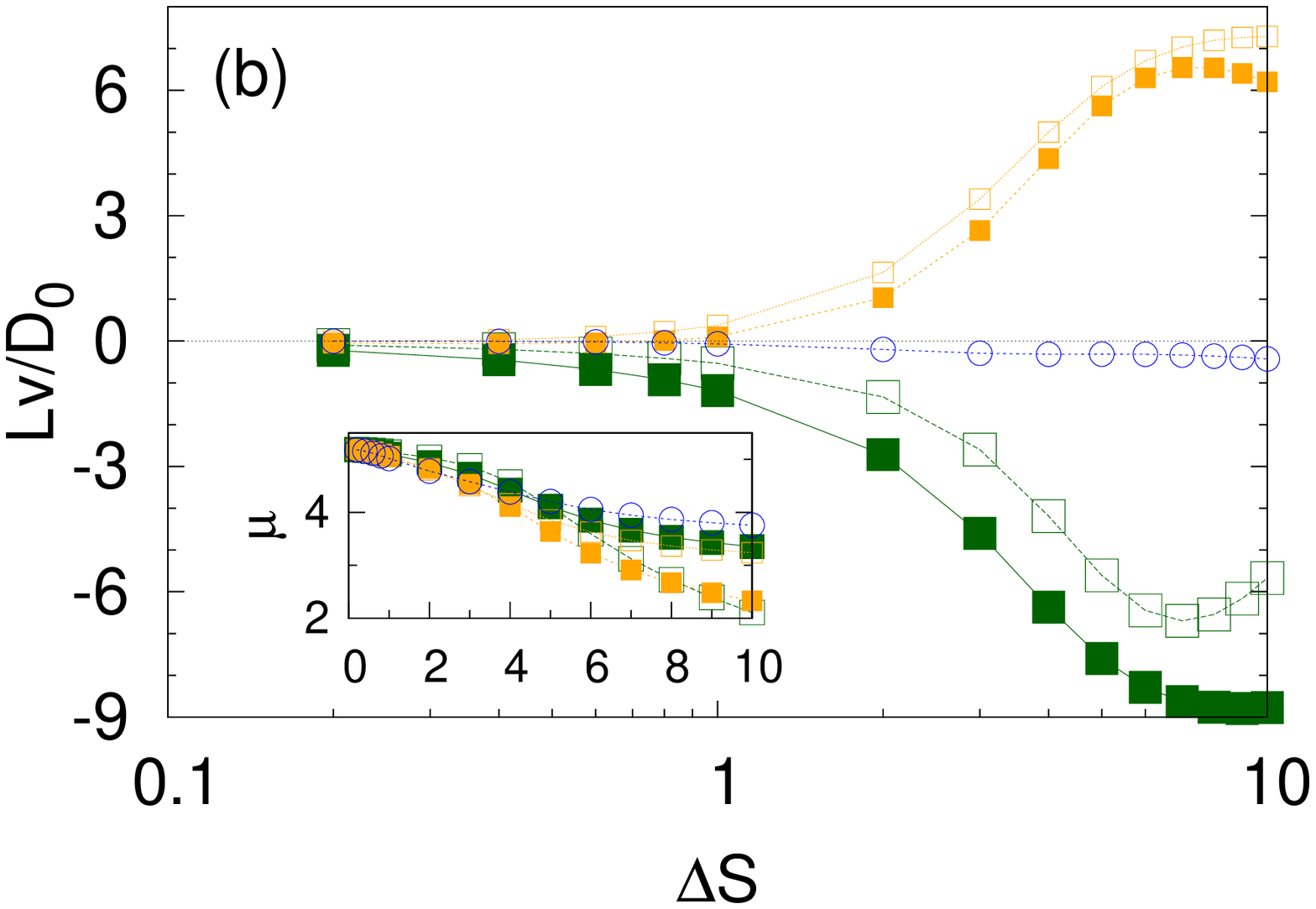}
\includegraphics[scale=0.23]{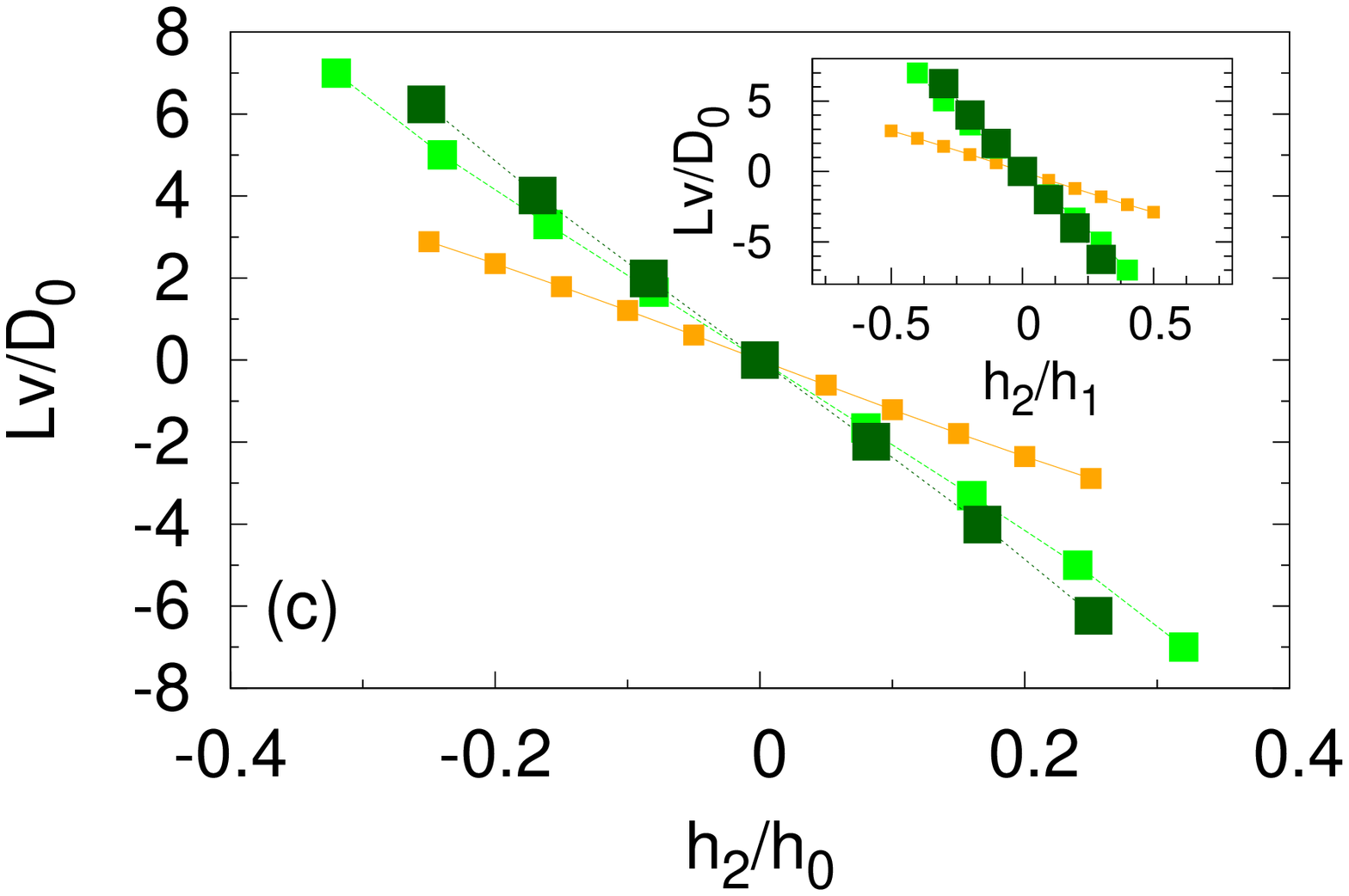}
\caption{Rectification of a Brownian motor moving due to a symmetric flashing ratchet in an asymmetric channel, for $h_2/h_1=0.25$. (a): Particle velocity, in units of $D_{0}/L$, $D_0=D_0(R=1)$, as a function of the phase shift $\phi_0$  for different values of the parameter $\Delta S=0.84,2.19,2.94$ (the larger the symbol   size, the larger $\Delta S$), with  $\mathcal{V}_1=0.2$ and $Q=2$. Inset: $\mu$ as a function of $\phi_0$ for the same parameters. (b): Particle velocity as a function of $\Delta S$, varied   increasing $h_1$ (with $R=1,h_0=1.25$) at constant ratio $h_2/h_1=0.1$ (open points) and $h_2/h_1=0.25$ (solid points), for $\phi_0=0.1,0.5$ and $\mathcal{V}_1=0.2,Q=2$ (the larger the symbol   size, the larger $\phi_0$) . Cyan open circles represent the average velocity obtained by a uniform distribution of $\phi_0$ as a function of $\Delta S$. Inset: $\mu$ as a function of $\Delta S$ for the same parameters. (c): Particle velocity as a function of the channel asymmetry parameter $h_2$, with $\phi_
0=0$ and $\Delta S=1.09,2.19$ (the larger the symbol   size, the larger $\Delta S$), for $R=1,h_0=1.25,\mathcal{V}_1=0.2,Q=2$. Inset: $\mu$ as a function of $h_2$ for the same parameters.
}
\label{fig-reimann-asimm-simm}
\end{figure}

\emph{Flashing ratchet}. 
Fig.~\ref{fig-reimann-asimm-simm}.a shows that the channel asymmetry leads to asymmetric net particle current as a function of the phase shift, $\phi_0$. Non-vanishing average velocities develop even  when the channel and the ratchet are in  registry, $\phi_0=0,1/2,1$. Hence now a mean, non-vanishing velocity can persist for a uniform distribution of $\phi_0$ as shown in the inset of Fig.~\ref{fig-reimann-asimm-simm}.b. The average particle velocity in the case of a broader distribution of $\phi_0$ is quite reduced with respect to the values obtained for a single fixed value of $\phi_0$, see fig.\ref{fig-reimann-asimm-simm}.b. Moreover, the average velocity in the former case in not significantly affected by the channel corrugation, while the dimensionless mobility, $\mu$,  takes values comparable to those obtained in the case of a fixed $\phi_0$. The channel asymmetry also enhances the rectifying velocity magnitude, which is  almost two fold larger than the one obtained for the symmetric channel (Fig.~\ref{
fig-reimann-simm-simm}.a). The net velocity also shows a strong dependence on the channel asymmetry, $\Delta S$. 
 The  insets of Fig.~\ref{fig-reimann-asimm-simm} show that $\mu>1$ for all  regimes considered, underlying the strong cooperative regime between the symmetric ratchet and the geometric confinement. Finally, we register a linear dependence of the particle velocity upon increasing $h_2$ at fixed $h_1$. As also shown in fig.~\ref{fig-reimann-asimm-simm}.b, the slope of the linear relation between $h_2$ and $\bar{v}$ depends on $\Delta S$: increasing the entropic barrier leads to a steeper slope.

\begin{figure}
 \includegraphics[scale=0.23]{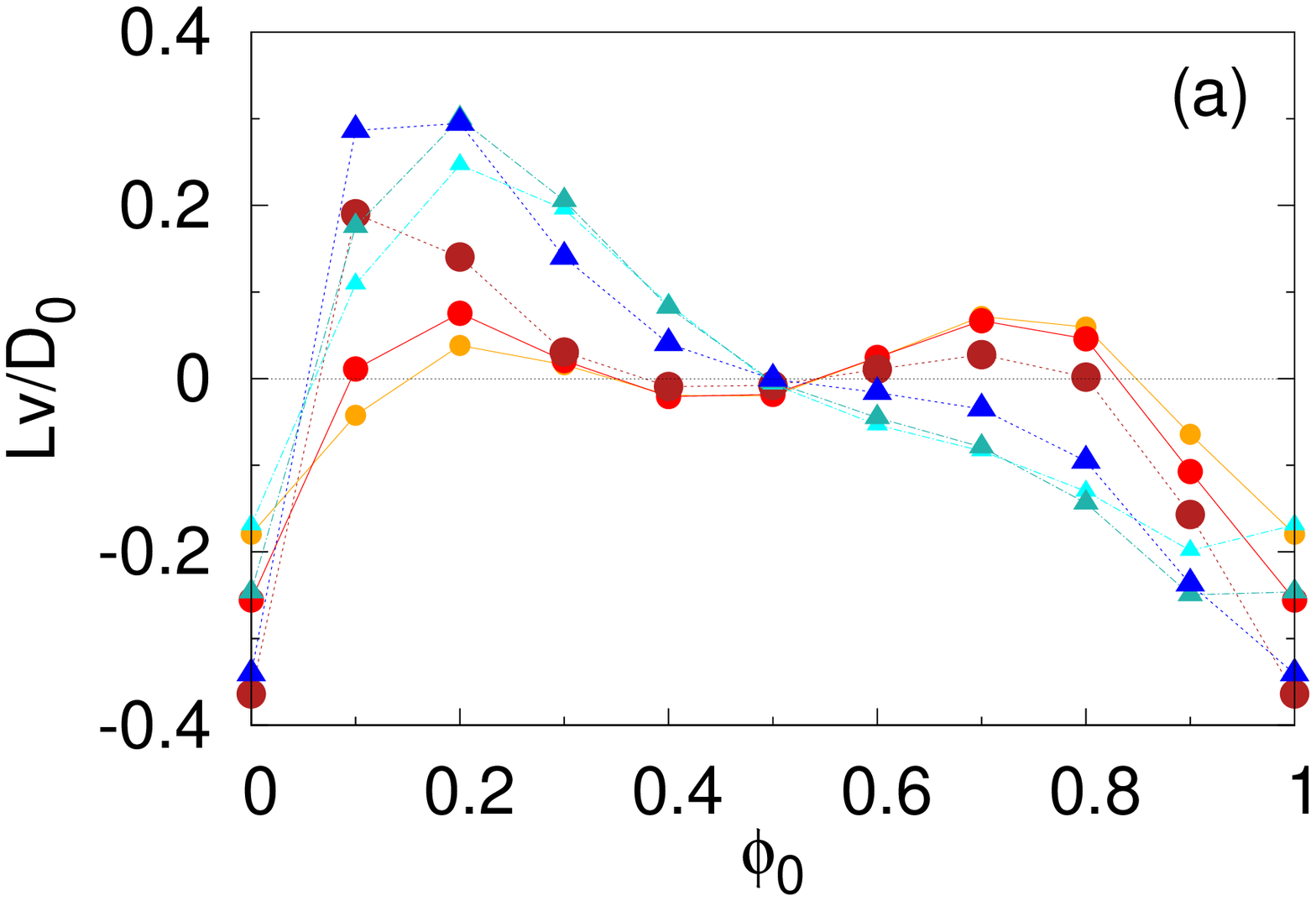}\includegraphics[scale=0.23]{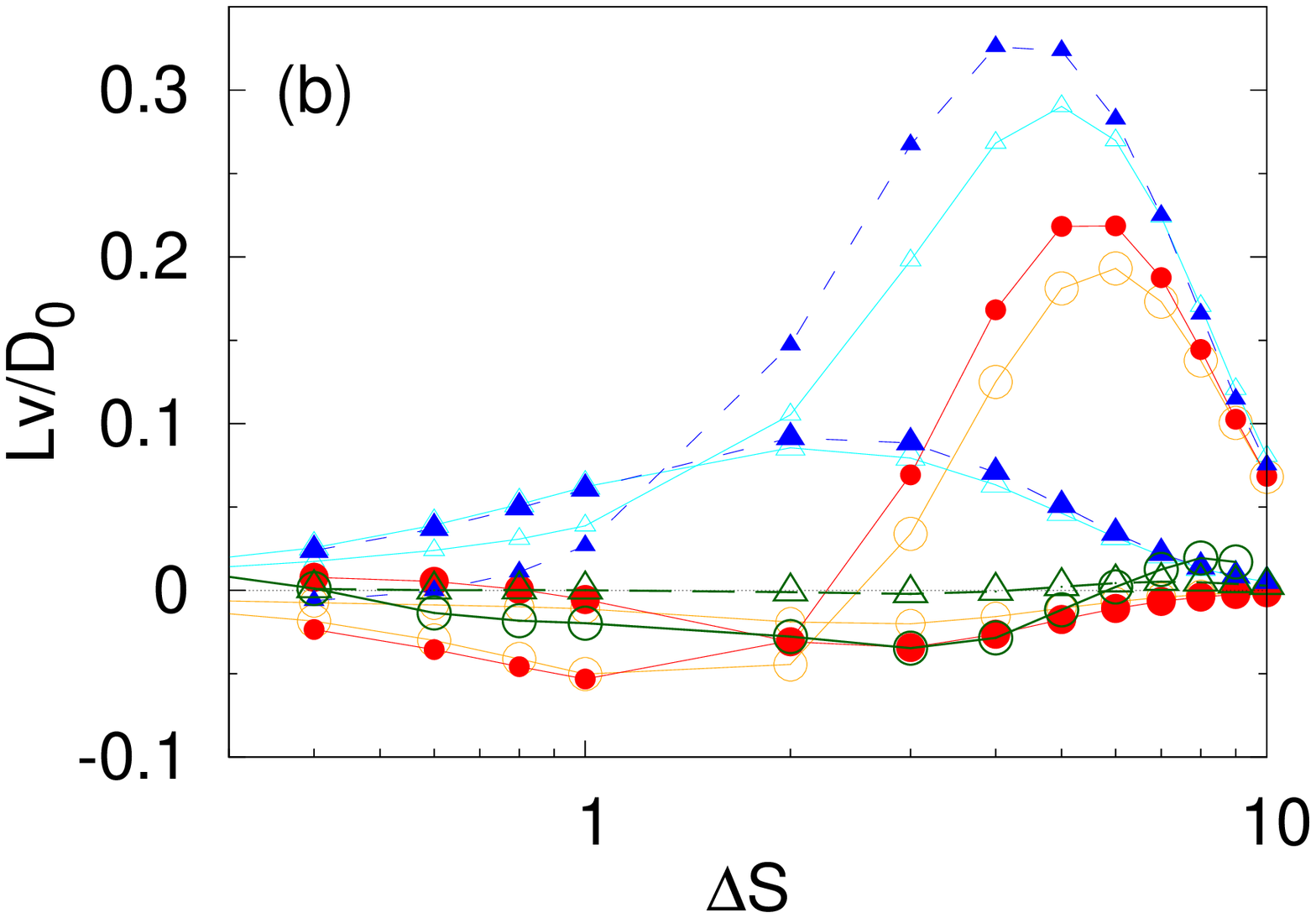}
\includegraphics[scale=0.23]{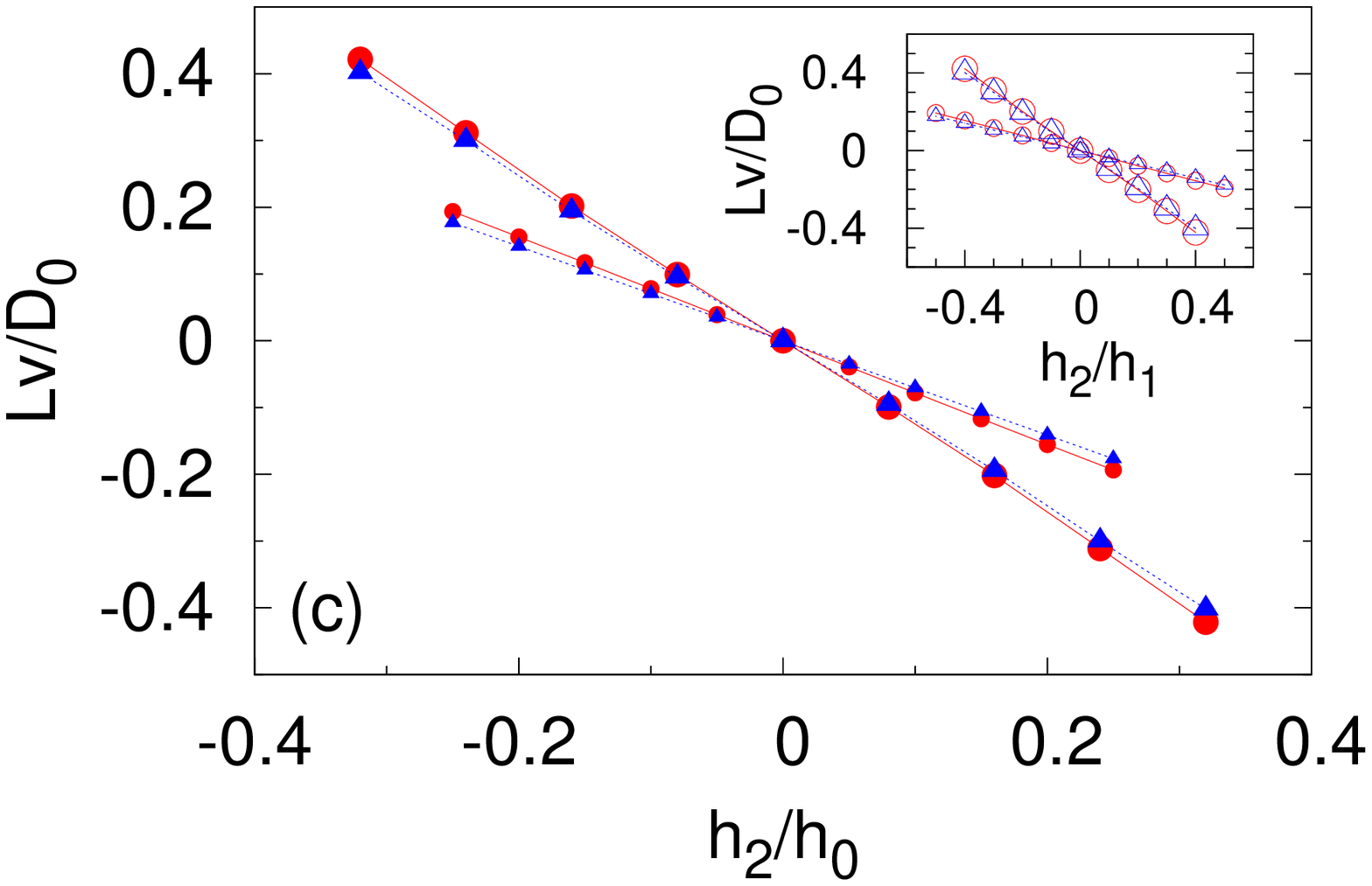}
\caption{Rectification of a processive (circles), non-processive (triangles) Brownian motor moving according  to the two state model in symmetric ratchet and asymmetric channel, for $h_2/h_1=0.25$. (a): Particle velocity, in units of $D_{0}/L$, $D_0=D_0(R=1)$, as a function of the phase shift $\phi_0$  for different values of the parameter $\Delta S=1.73,2.19,2.94$ (the larger the symbol   size, the larger $\Delta S$), with  $\mathcal{V}_1=0.2$ and $\omega_{2,1}/\omega_{1,2}=0.01$. (b): Processive (circles), non-processive (triangles) Brownian motor velocity, in units of $D_{0}/L$, as a function of $\Delta S$ as a function of particle radius $R$ (solid lines, with $h_0=1.25,h_1=0.2$) or $h_1$ (open points, with $R=1,h_0=1.25$) for $\phi_0=0.1,0.4$ (the larger the symbol   size, the larger $\phi_0$) for $\mathcal{V}_1=1$ and $\omega_{2,1}/\omega_{1,2}=0.01$. Green open circles (triangles)  represent the average velocity of processive (non-processive) motors obtained by a uniform distribution of $\phi_0$ as a 
function of $\Delta S$. (c): processive (circles), non-processive (triangles) Brownian motor velocity as a function of $h_2$, being $\phi_0=0$, with $h_0=1.25,\mathcal{V}_1=0.2,\omega_{2,1}/\omega_{1,2}=0.01$. The larger the symbol   size, the larger the value of $h_1$ ($h_1=0.125,0.2$). 
}
\label{fig-twostate-asimm-simm}
\end{figure}

\emph{Two-state model.} As for the flashing ratchet, the channel asymmetry leads to an asymmetric velocity profile upon variation of $\phi_0$, as shown in Fig.~\ref{fig-twostate-asimm-simm}.a. In particular, we notice that for $\phi_0=0$ the net velocity of processive or non-processive motors are very similar as $\Delta S$ varies (fig.~\ref{fig-twostate-simm-simm}.a). The asymmetric channel profile leads to an overall non vanishing flux when averaged over the, equally weighted, values of $\phi_0$. Hence, as for the flashing ratchet, we expect the channel asymmetry to provide the onset of net currents even in the case of a broader distribution of phase shifts $\phi_0$. 
The dependence of the particle velocity on $\Delta S$, shown in fig.~\ref{fig-twostate-asimm-simm}.b, is similar to the one obtained for the symmetric channel, fig.~\ref{fig-twostate-simm-simm}.b, where an optimal value of $\Delta S$ is observed for both processive and non processive motors.  As in the symmetric configuration, the dependence of motors' velocity on the different geometric parameters ($h_1$ and $R$) is quite well captured by the entropic barrier $\Delta S$,  although a separate sensitivity on $h_1$ and $R$ persists for both processive and non-processive motors. Particle current depends linearly on $h_2$, as shown in fig.~\ref{fig-twostate-asimm-simm}.c. increasing the slope for larger values of  $\Delta S$.

\section{VI. Asymmetric potential and symmetric channel}
An asymmetric ratchet potential, $\lambda \ne 0$, in the presence of a  symmetric periodic channel, $h_2=0$,  allows us to address the impact that an inhomogeneous environment has on an intrinsically rectifying Brownian ratchet. In particular,  cooperative rectification modulates the particle velocity allowing for the emergence of effective particle  fluxes opposing the direction of motion of the intrinsic Brownian ratchet.

\begin{figure}
 \includegraphics[scale=0.23]{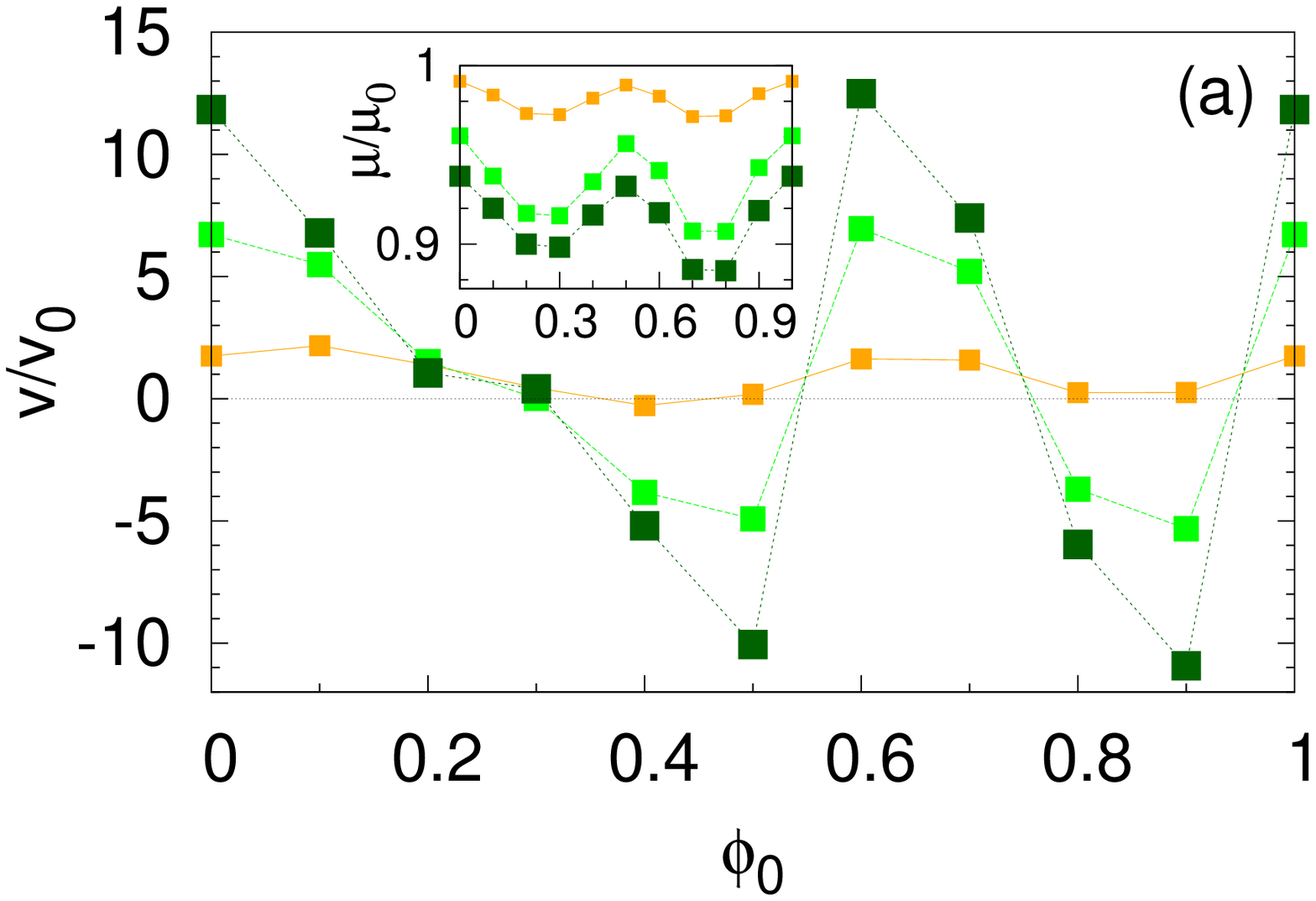}\includegraphics[scale=0.23]{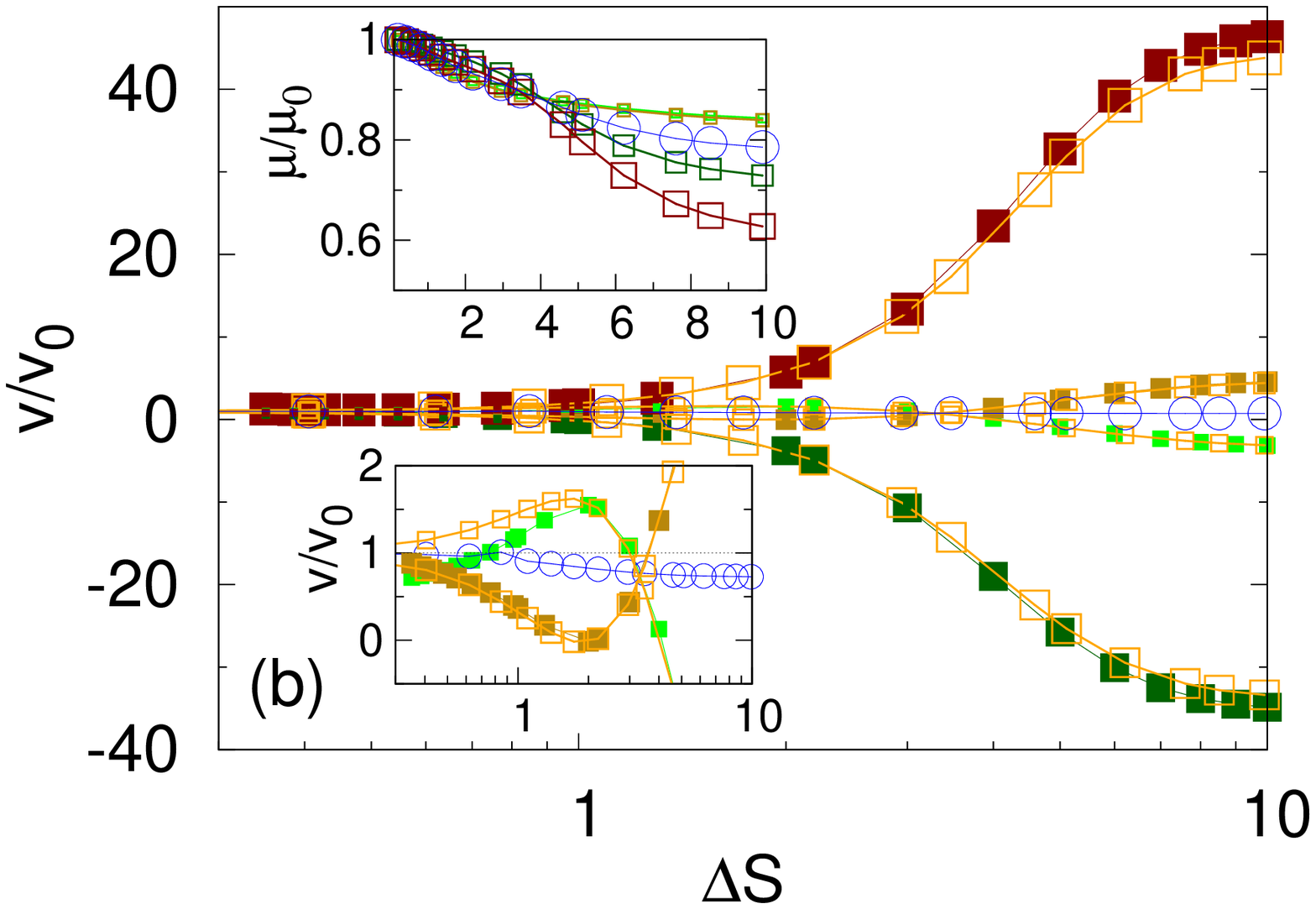}
\includegraphics[scale=0.23]{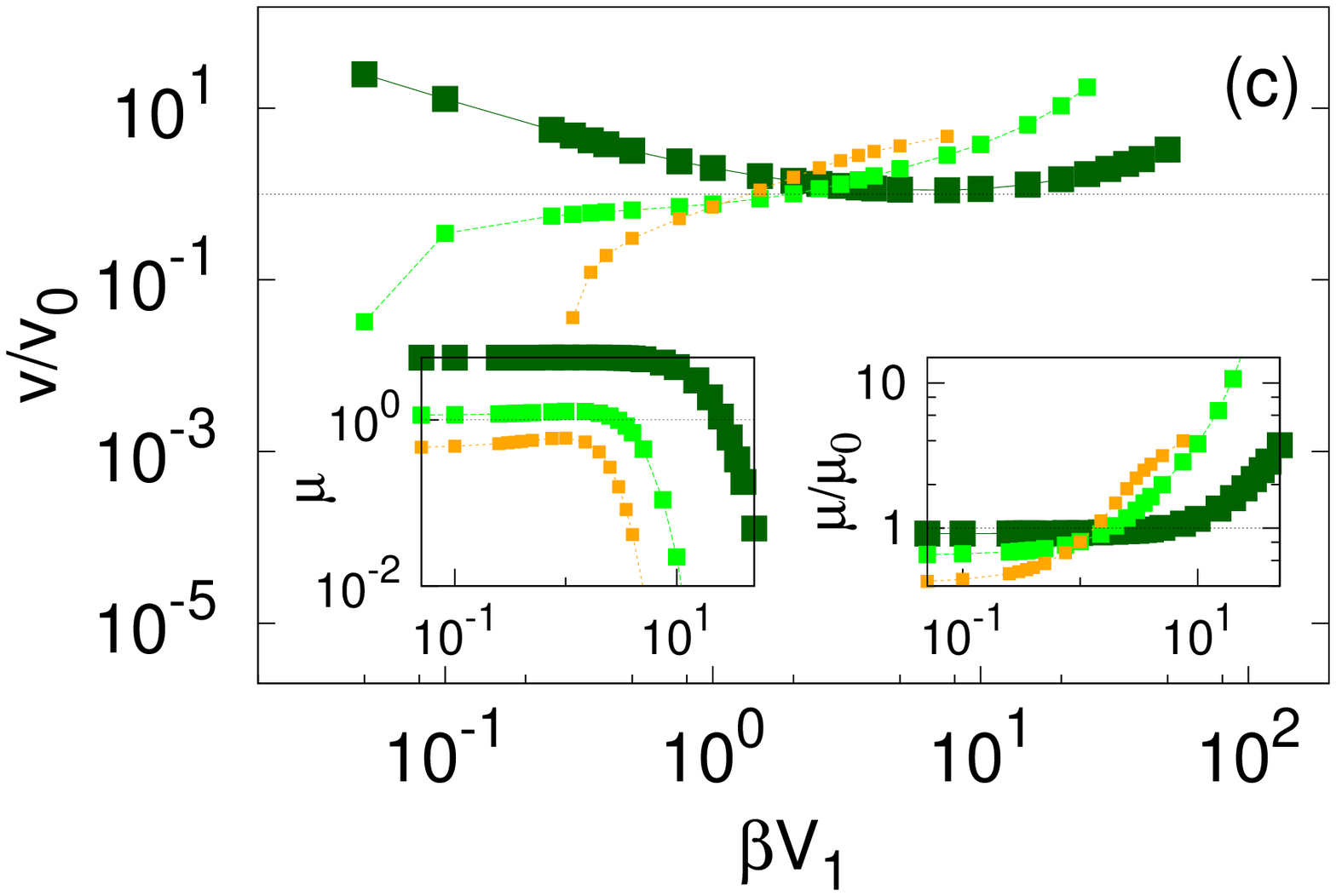}\includegraphics[scale=0.23]{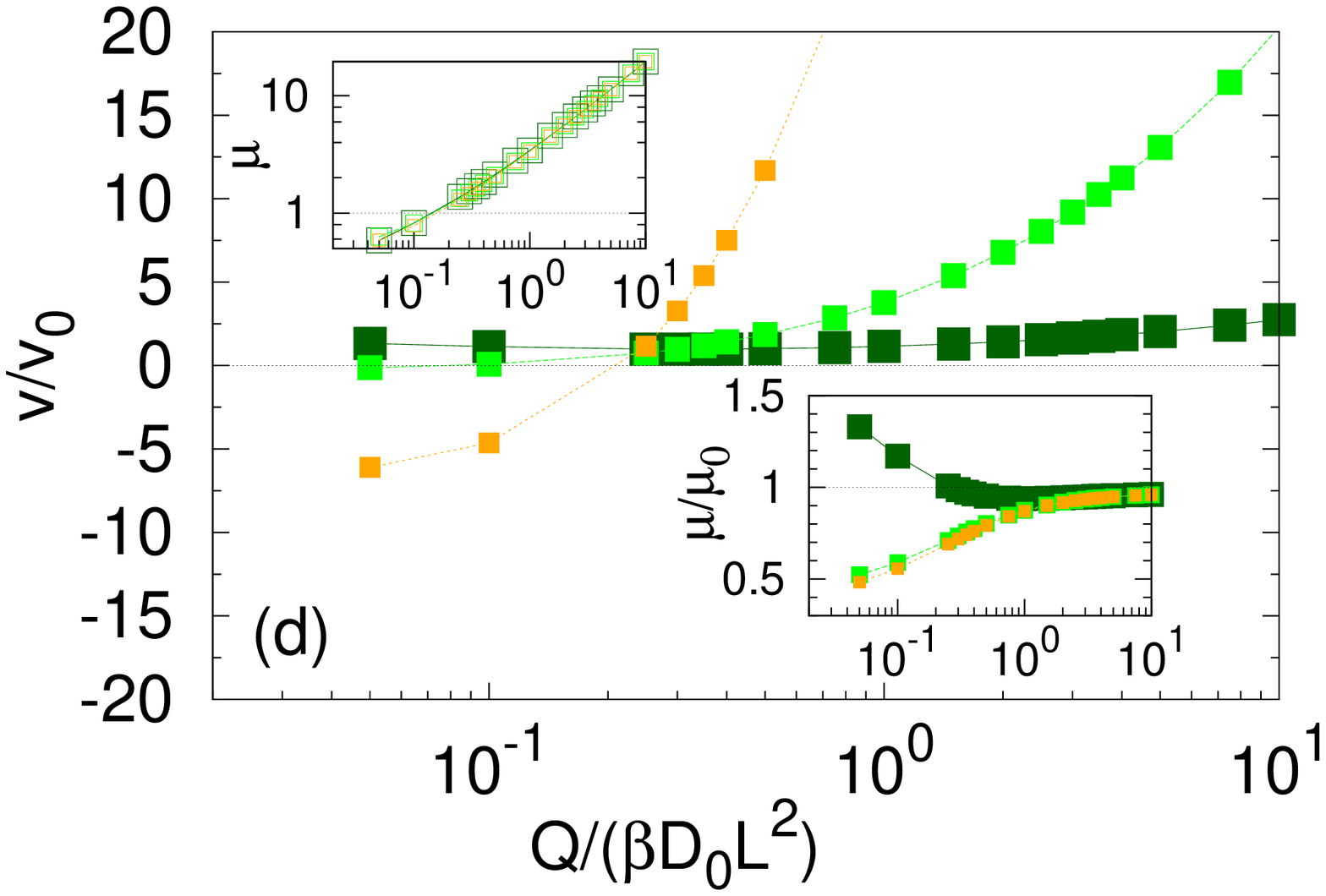}
\caption{Rectification of a Brownian motor moving due to a asymmetric flashing ratchet in symmetric channel. (a): particle velocity, in units of  $v_{0}$, for $\Delta S=0$, as a function of the phase shift $\phi_0$  for different values of the parameter $\Delta S=0.84,2.19,2.94$ (the larger the symbol   size, the larger $\Delta S$), for  $\mathcal{V}_1=0.2$ and $Q=2$. Inset: $\mu$, in units of the dimensionless mobility, $\mu_0$,  as a function of $\phi_0$ for the same parameters. (b): particle velocity as a function of $\Delta S$ as a function  of particle radius $R$ (solid lines, with $h_1=1.25,h_2=0.2$), $h_1$ (solid points, with $R=1,h_2=0.2$) or $h_2$ (open points, with $R=1,h_1=1.25$) for $\phi_0=0.2,0.3,0.5,0.6$ and $\mathcal{V}_1=0.2,Q=2$ (the larger the symbol   size, the larger $\phi_0$). Cyan open circles represent the average velocity obtained by a uniform distribution of $\phi_0$ as a function of $\Delta S$. Inset: $\mu/\mu_0$ as a function of $\phi_0$ for the same parameters. (c): particle 
velocity as a function of the ratchet potential amplitude, $\mathcal{V}_1$, for $Q=0.02,0.2,2$ (the larger the symbol   size, the larger $Q$), for $\Delta S=2.94,\phi_0=0.1$. Insets: $\mu$ and $\mu/\mu_0$ as a function of $\phi_0$ for the same parameters. (d): particle velocity as a function of $Q$. Squares for $\mathcal{V}_1=0.1,1,10$ and $\Delta S=2.94,\phi_0=0.1$, (the larger the symbol   size, the larger $\mathcal{V}_1$); triangles: $Q=1$, $\Delta S=1.1$, $\phi_0=0.5$ and $\mathcal{V}_1=1$. Insets: $\mu$ and $\mu/\mu_0$ as a function of $\phi_0$ for the same parameters.
}
\label{fig-reimann-simm-asimm}
\end{figure}
\emph{Flashing ratchet}. 
Fig.~\ref{fig-reimann-simm-asimm}.a shows that the intrinsic Brownian ratchet net velocity, $v_0$, is strongly  modulated by the channel corrugation. CBR in this case can exhibit both regimes where the average  velocities   exceed $v_0$, showing strong velocity enhancements, as well as conditions where  the velocity changes sign, indicating confinement-induced flow reversal.  In the latter case, particles moving against the direction imposed by the ratchet can display speeds larger than $v_0$.
As in the case of symmetric ratchet, these effects are magnified when rising the entropy barrier $\Delta S$, as shown in Fig.~\ref{fig-reimann-simm-asimm}.b.  In the presence of flux reversal, geometrical confinement leads to a mechanism for particle separation based on their size because  $\Delta S$ depends both  on the channel geometry and the particle size. As shown in fig.~\ref{fig-reimann-simm-asimm}.b, modulating particles size one can control and  switch their velocities, offering new venues to manipulate particles and even trap them. The average particle current obtained in a disordered channel, i.e. with a uniform distribution of $\phi_0$, is not much affected by the geometrical constraint. 

The absolute value of particle current (not shown) is also very sensitive to   $\mathcal{V}_1$ and $Q$ , analogously to the  results obtained for the symmetric channel (Fig.~\ref{fig-reimann-simm-simm}.c-\ref{fig-reimann-simm-simm}.d). Figs.~\ref{fig-reimann-simm-asimm}.c-\ref{fig-reimann-simm-asimm}.d display strong enhancements of the net particle velocity, up to two orders of magnitude larger than  $v_0$. These large enhancements are observed when  $\mathcal{V}_1/\Delta S \ll 1$, indicating that  cooperativity  relies mostly on the interplay between the geometrical confinement and the position-dependent noise amplitude rather than on the asymmetric potential $V(x)$ itself. 

Since the Brownian ratchet is characterized by an intrinsic rectifying velocity, $v_0$, it is useful to study
the ratio $\mu/\mu_0$ in order to quantify the relative  variation in the mobility of a CBR due to the geometrical constraints. In Figs.~\ref{fig-reimann-simm-asimm}.a-\ref{fig-reimann-simm-asimm}.b $\mu_0=6.1$, hence the system takes advantage of the $x$-dependent free energy gradient induced by   the intrinsic ratchet mechanism. The dependence of $\mu$ on  $\mathcal{V}_1$ and $Q$ is quite similar to the one observed in symmetric channels (Fig.~\ref{fig-reimann-simm-simm}): larger values of the mobility are registered for small and moderate values of $\mathcal{V}_1$ and mild and large values of $Q$, while for larger values of $\mathcal{V}_1$ $\mu$ drops. On the contrary, $\mu/\mu_0$ increases monotonously with $\mathcal{V}_1$ irrespectively of $Q$, while the dependence on $Q$ at fixed $\mathcal{V}$ is more involved as shown in fig.~\ref{fig-reimann-simm-asimm}.d.

\begin{figure}
 \includegraphics[scale=0.23]{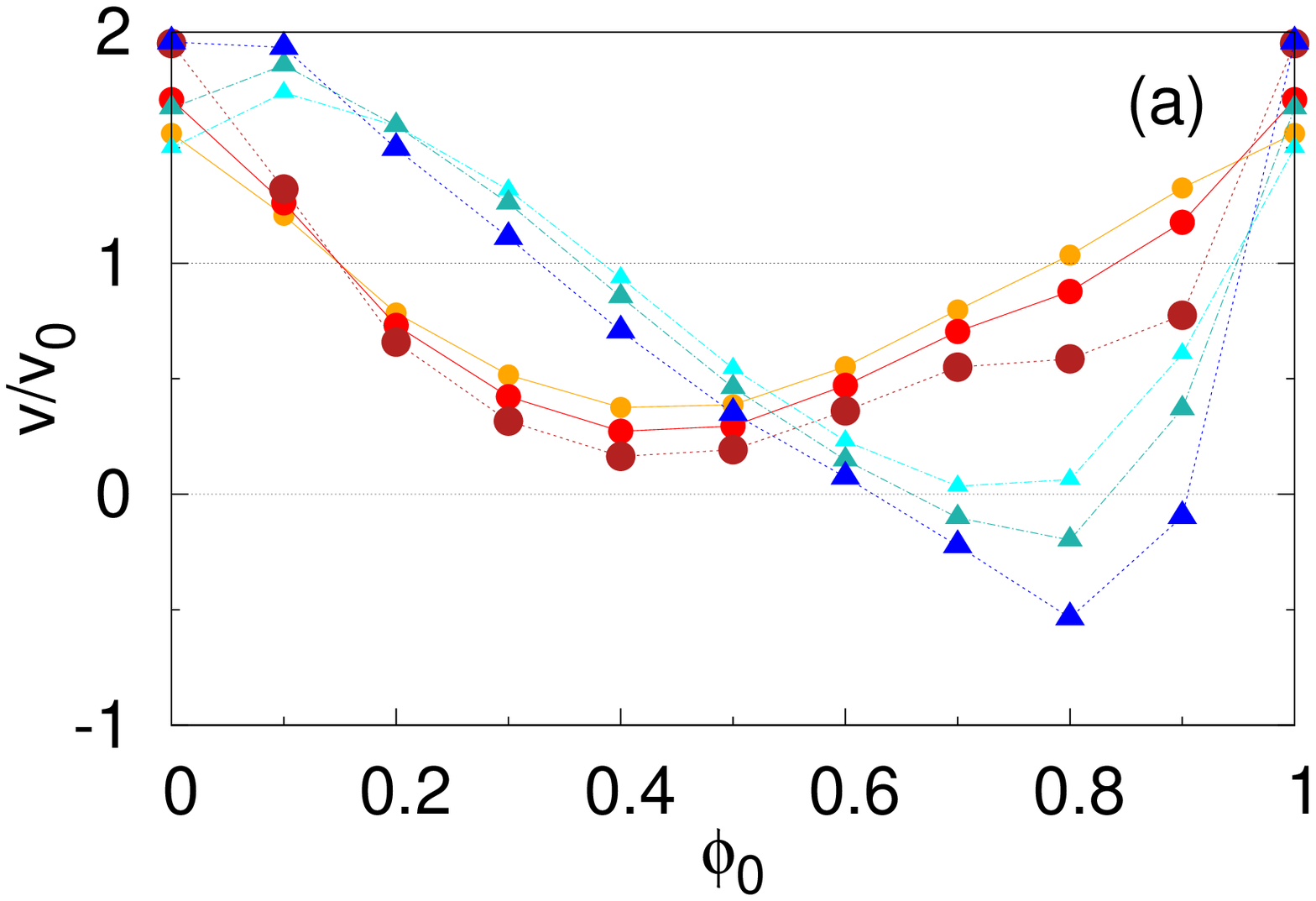}\includegraphics[scale=0.23]{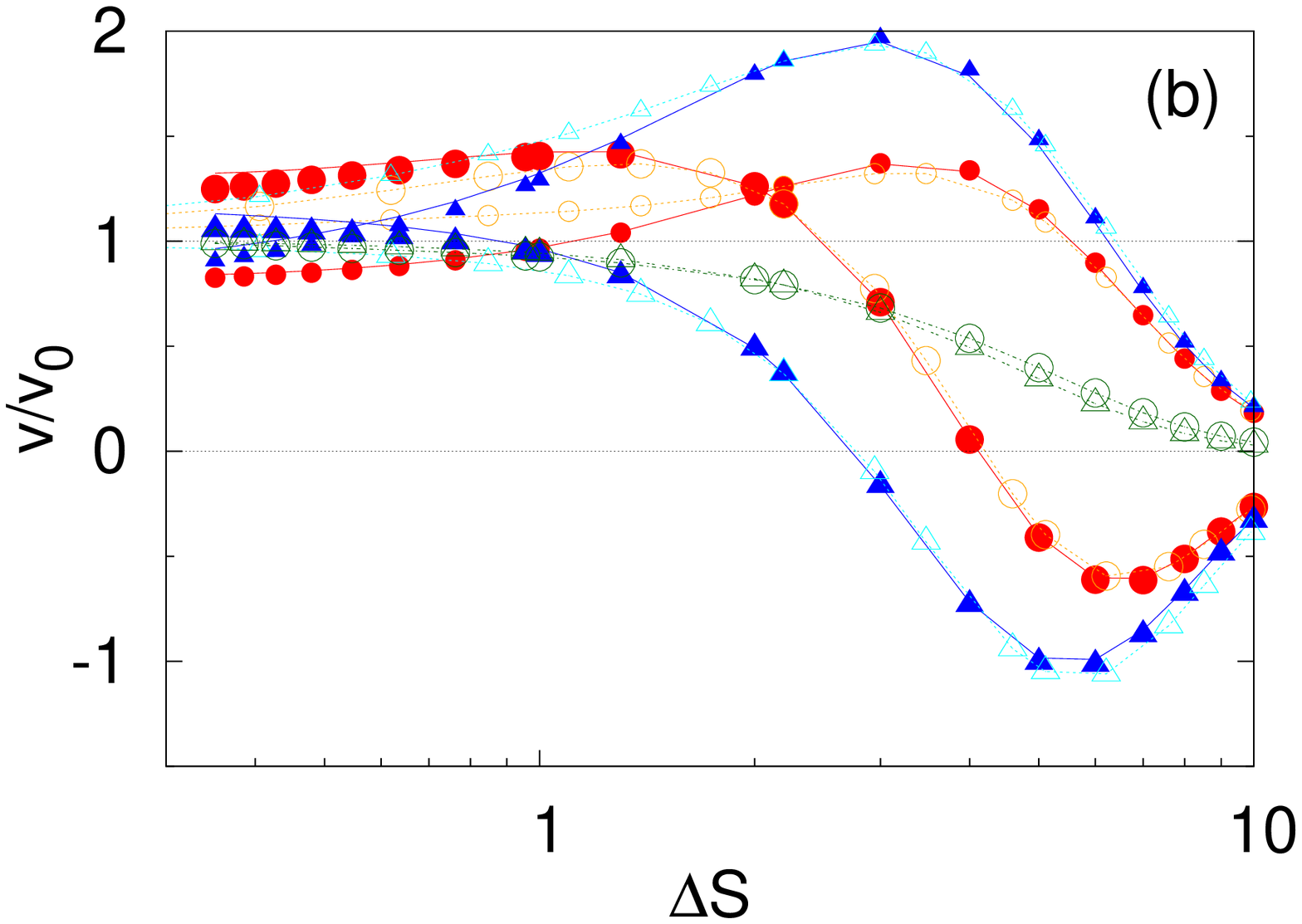}
\includegraphics[scale=0.23]{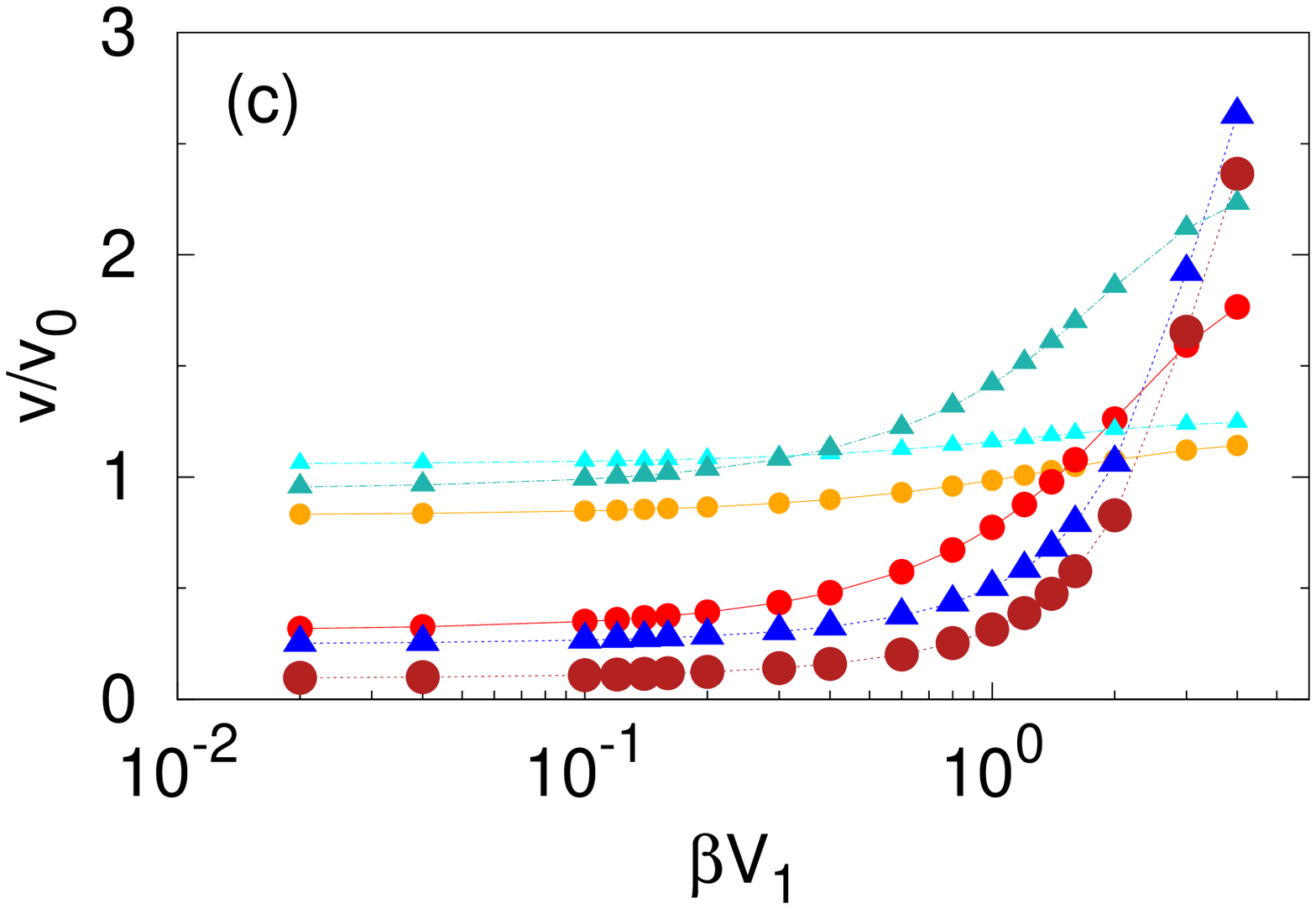}\includegraphics[scale=0.23]{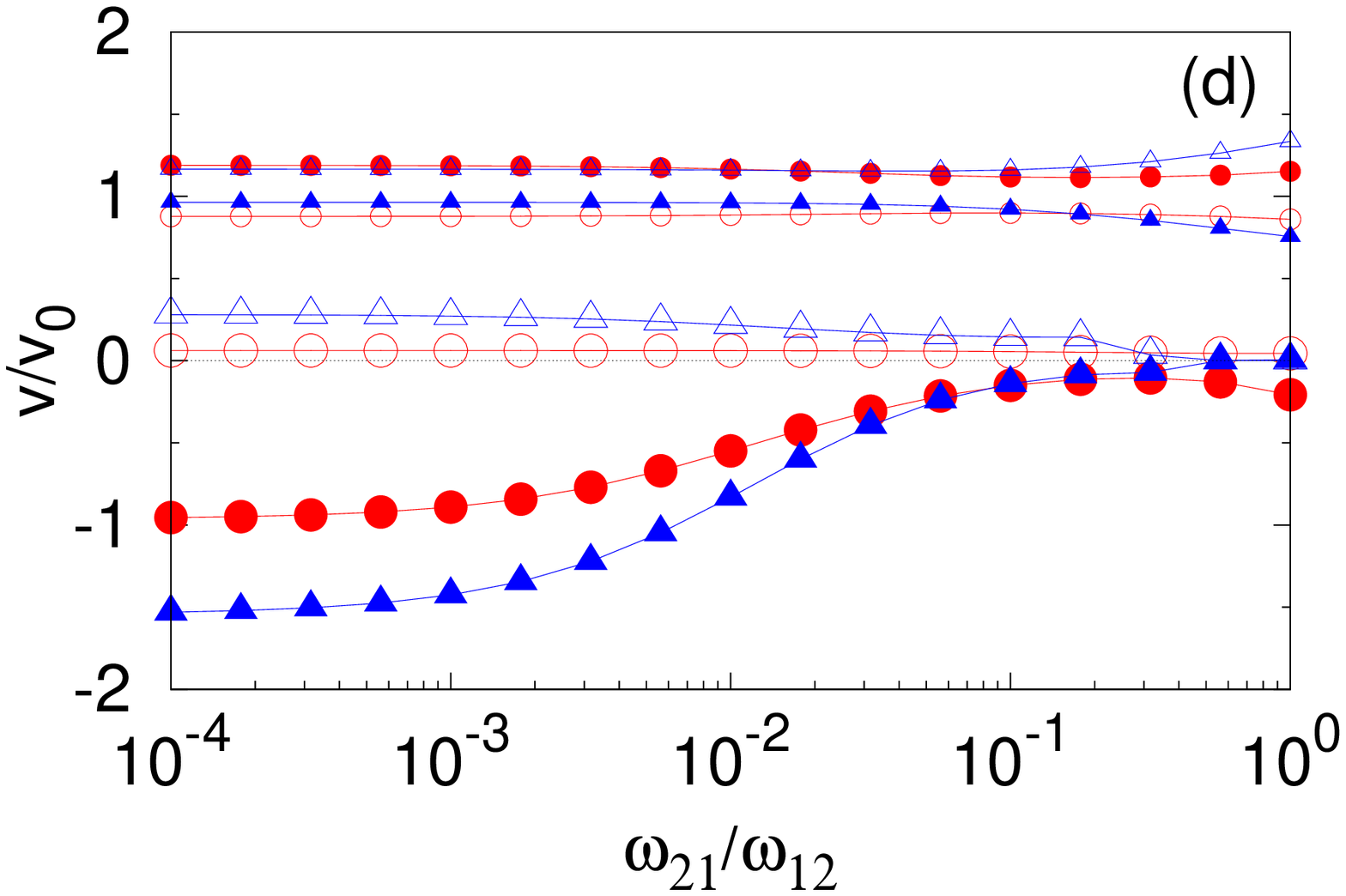}
\caption{Rectification of a processive (circles), non-processive (triangles) Brownian motor moving due to the two state model in symmetric channel. (a): particle velocity, in units of $v_{0}$ as a function of the phase shift $\phi_0$  for different values of the parameter $\Delta S=1.73,2.19,2.94$ (the larger the symbol   size, the larger $\Delta S$), for  $\Delta V_1=0.2$ and $\omega_{2,1}/\omega_{1,2}=0.01$. (b): processive (circles), non-processive (triangles) Brownian motor velocity, in units of $D_{0}/L$, as a function of $\Delta S$ and particle radius $R$ (solid lines, with $h_0=1.25,h_1=0.2$), $h_0$ (solid points, with $R=1,h_1=0.2$) or $h_1$ (open points, with $R=1,h_0=1.25$) for $\phi_0=0.1,0.9$ (the larger the symbol   size, the larger $\phi_0$)  for $\mathcal{V}_1=1$  and $\omega_{2,1}/\omega_{1,2}=0.01$. Green open circles (triangles)  represent the average velocity of processive (non-processive) motors obtained by a uniform distribution of $\phi_0$ as a function of $\Delta S$. (c): processive (
circles), non-processive (triangles) Brownian motor velocity as a function of the ratchet potential amplitude $\mathcal{V}_1$ for $\Delta S=0.4,2,7.6$ (the larger the symbol   size, the larger $\Delta S$)  with $\omega_{2,1}/\omega_{1,2}=0.01$. (d): processive (circles), non-processive (triangles) Brownian motor velocity as a function of $\omega_{1,2}/\omega_{2,1}$ for $\phi_0=0.1,0.3$, open (solid) points and $\Delta S=0.4,7.6$(the larger the symbol   size, the larger $\Delta S$) for $\mathcal{V}_1=10$.
}
\label{fig-twostate-simm-asimm}
\end{figure}
\emph{Two state model}. 
Fig.~\ref{fig-twostate-simm-asimm}.a displays the net velocity as a function of the dephasing between the geometric confinement and the underlying ratchet potential. Cooperative rectification now shows a strong dependence on the phase shift, $\phi_0$, leading to  large velocity amplification and also to flux reversal, a feature that was not  possible for symmetric channels.  In fact, both processive and non-processive motors  show velocity enhancement and reversal when varying the channel corrugation, $\Delta S$, as shown in Fig.~\ref{fig-twostate-simm-asimm}.b.   Hence, even symmetric channels offer the possibility to control molecular motor motion according to their size, allowing for segregation and particle trapping. 
 Interestingly, for asymmetric ratchets the entropic barrier $\Delta S$ captures even better the dynamics, as compared to the case of symmetric ratchets, and only at smaller $\Delta S$ the different behavior upon variation of $h_0,h_1$ and $R$ becomes appreciable.  Looking at the velocity dependence as a function of $\mathcal{V}_1$, shown in fig.~\ref{fig-twostate-simm-asimm}.c, we find a behavior similar to the one obtained for the symmetric ratchet. On the contrary, the velocity dependence upon variation of $\omega_{21}/\omega_{12}$, shown in fig.~\ref{fig-twostate-simm-asimm}.d, is very mild for smaller $\Delta S$ while for larger $\Delta S$ velocity inversion happens for smaller values of $\omega_{21}/\omega_{12}$.

\section{VII. Fully Asymmetric case}
When both the ratchet as well the channel left-right symmetry are broken, $\lambda \ne 0$ and $h_2 \ne 0$, all the features we have discussed previously are now present.  Rather than attempting a systematic analysis of the performance of CBR in this regime, that is very rich, we will point out the basic differences with the previous cases.

\begin{figure}
 \includegraphics[scale=0.23]{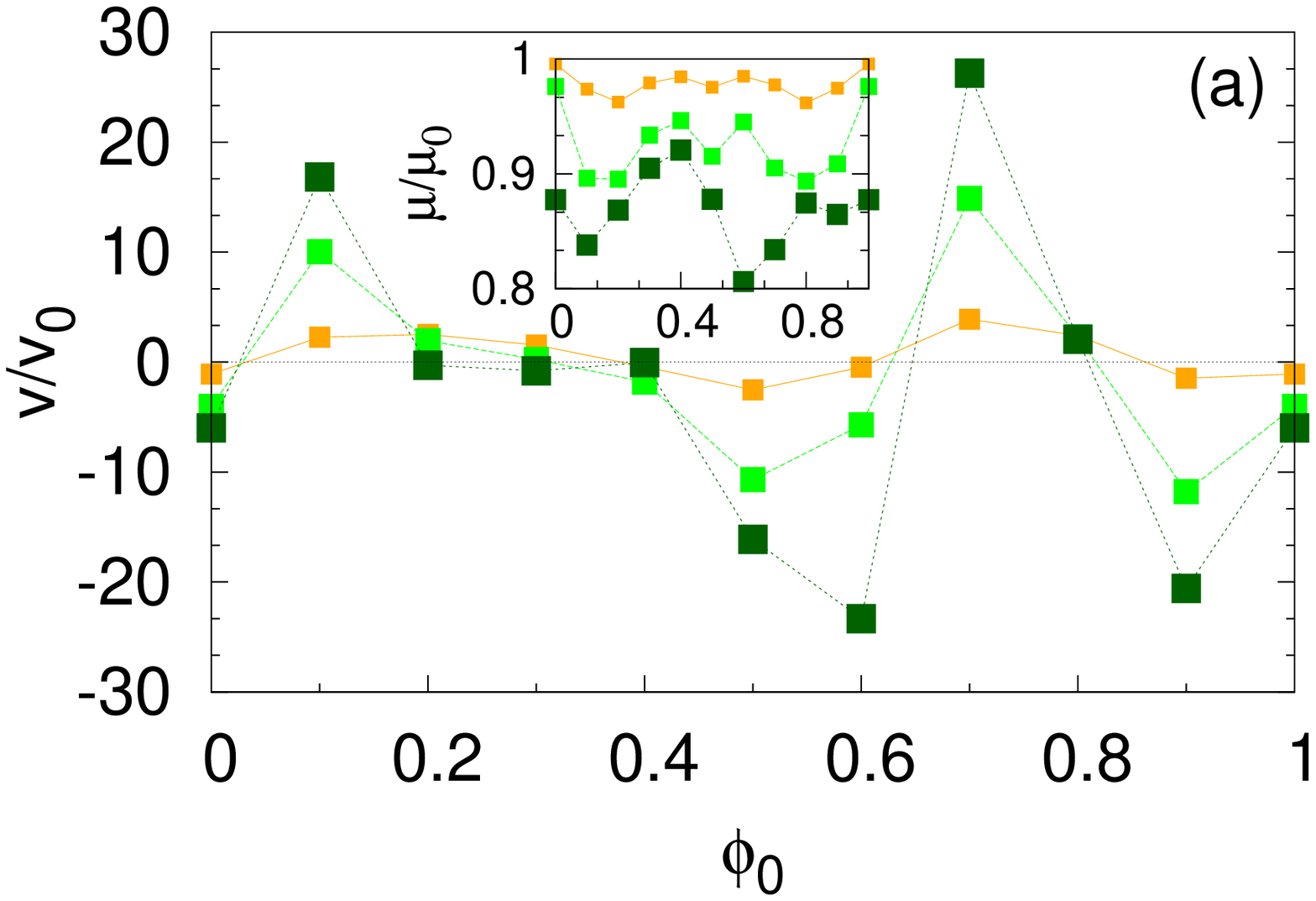}\includegraphics[scale=0.23]{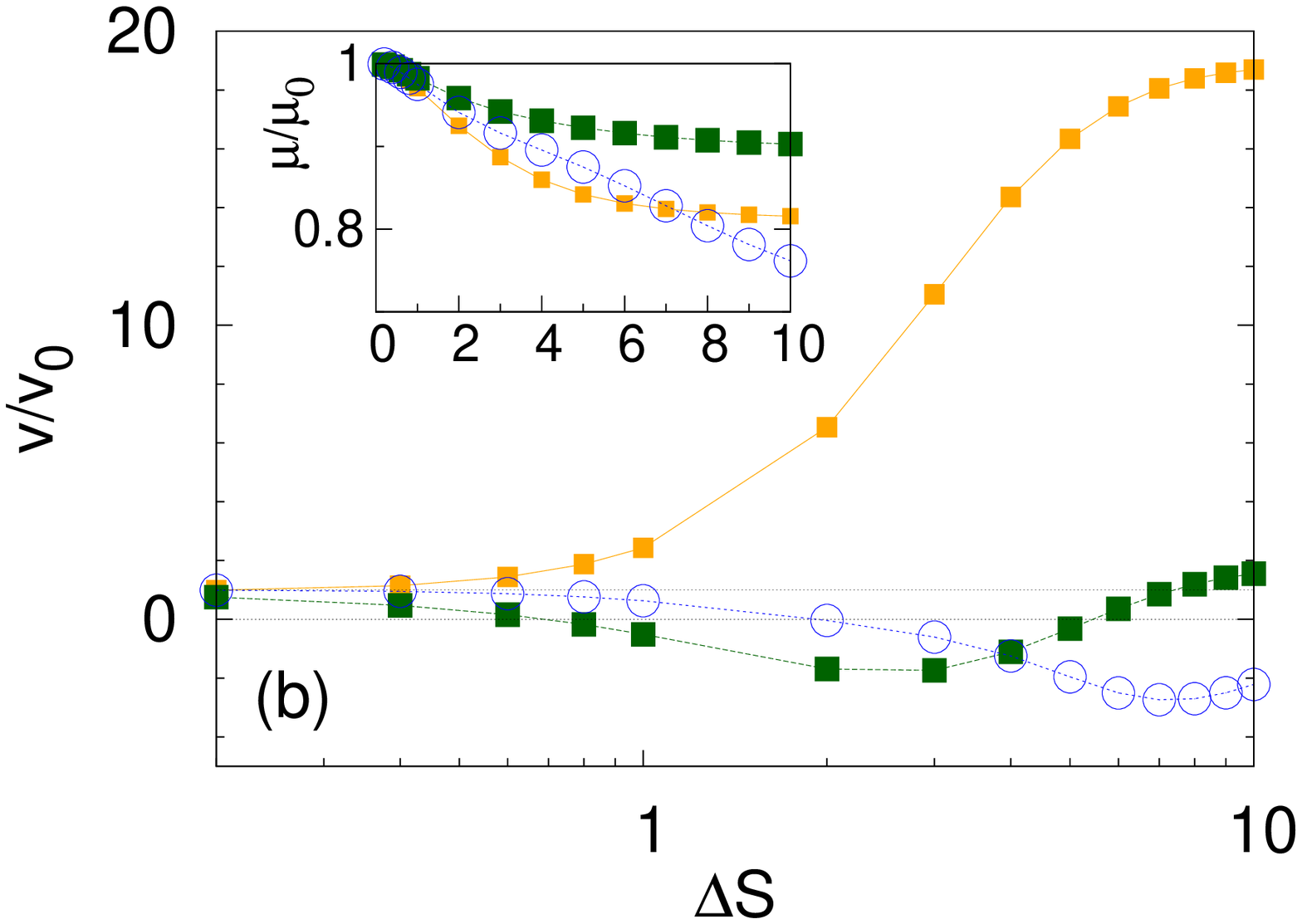}
\caption{Rectification of a Brownian motor moving according to an asymmetric flashing ratchet in an asymmetric channel. (a): particle velocity, in units of the velocity,$v_{0}$, provided by the ratchet for $\Delta S=0$, as a function of the phase shift $\phi_0$  for different values of the parameter $\Delta S=0.84,2.19,2.94$ (the larger the symbol   size, the larger $\Delta S$), for  $\mathcal{V}_1=0.2$, $h_2/h_1=0.25$ and $Q=2$. Inset: $\mu$, in units of the dimensionless mobility $\mu_0$ as a function of $\phi_0$ for the same parameters. (b): particle velocity as a function of $\Delta S$ and $h_1$ (with $R=1,h_0=0.25$) for $\phi_0=0.1,0.4$, $h_2/h_1=0.25$ and $\mathcal{V}_1=0.2,Q=2$  (the larger the symbol   size, the larger $\phi_0$). Cyan open circles represent the average velocity obtained by a uniform distribution of $\phi_0$ as a function of $\Delta S$. Inset: $\mu/\mu_0$ as a function of $\phi_0$ for the same parameters.
}
\label{fig-reimann-asimm-asimm}
\end{figure}
\emph{Flashing ratchet.} As shown in Fig.~\ref{fig-reimann-asimm-asimm}.a, the asymmetry in both  channel shape and ratchet potential lead to  non-intuitive velocity variations with  $\phi_0$; one can identify velocity enhancement/reduction as well as velocity inversion as a function of the off-registry angle. Moreover, different values of $\Delta S$ strongly modulate the velocity dependence on $\phi_0$ as shown in fig.~\ref{fig-reimann-asimm-asimm}.a. Looking at the dependence of the velocity on $\Delta S$, fig.~\ref{fig-reimann-asimm-asimm}.b, we  observe a strong non-monotonic response where the motor velocity, initially enhanced, is reduced by increasing $\Delta S$ until  it is inverted for larger values of $\Delta S$. As in the previous cases, we find a wide range of values of $\mu$ as well as of $\mu/\mu_0$  underlying that the sensitivity of  CBRs in this scenario as a function of the changes in the geometrical constraints and  ratchet parameters.

\begin{figure}
 \includegraphics[scale=0.23]{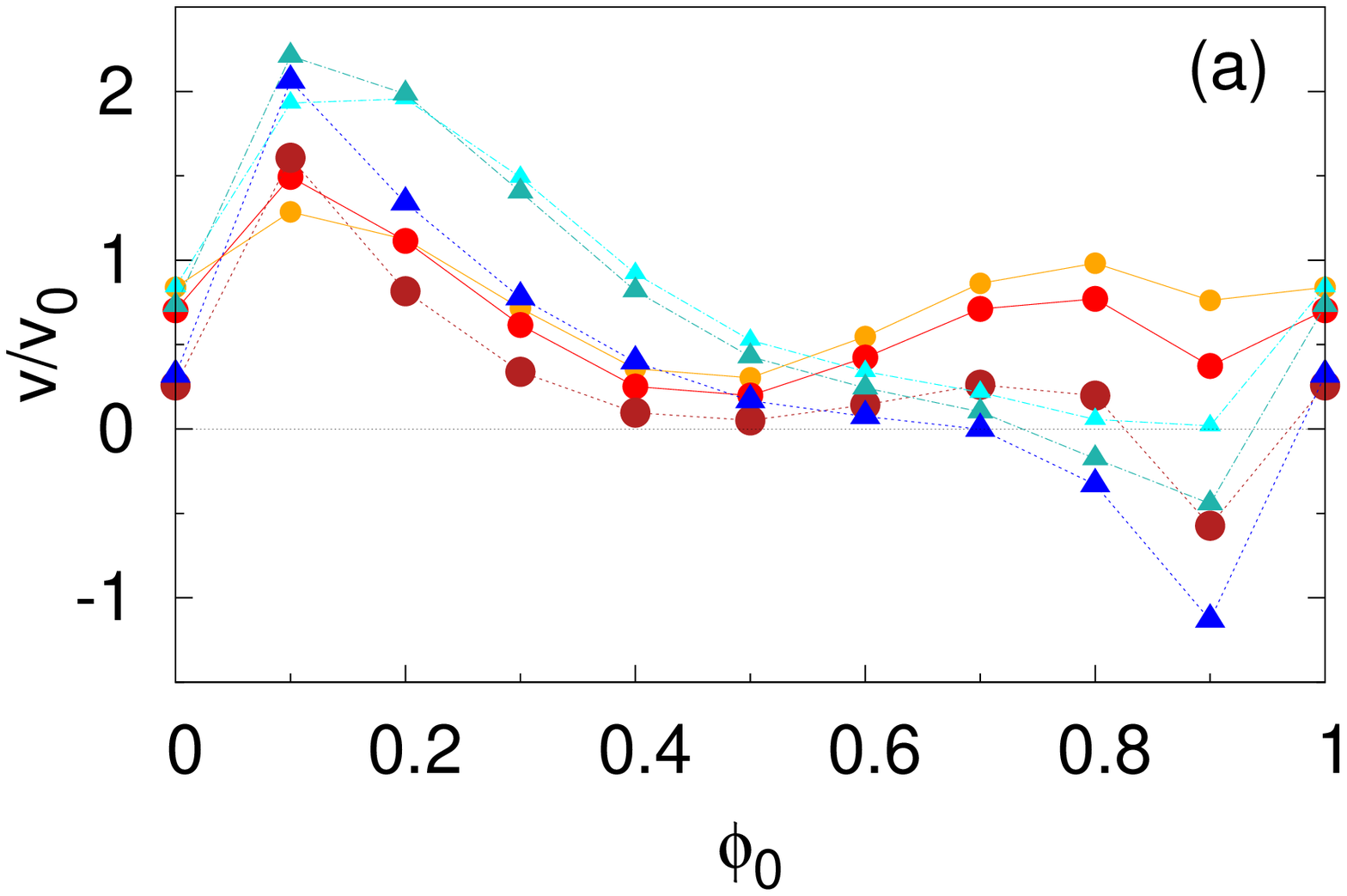}\includegraphics[scale=0.23]{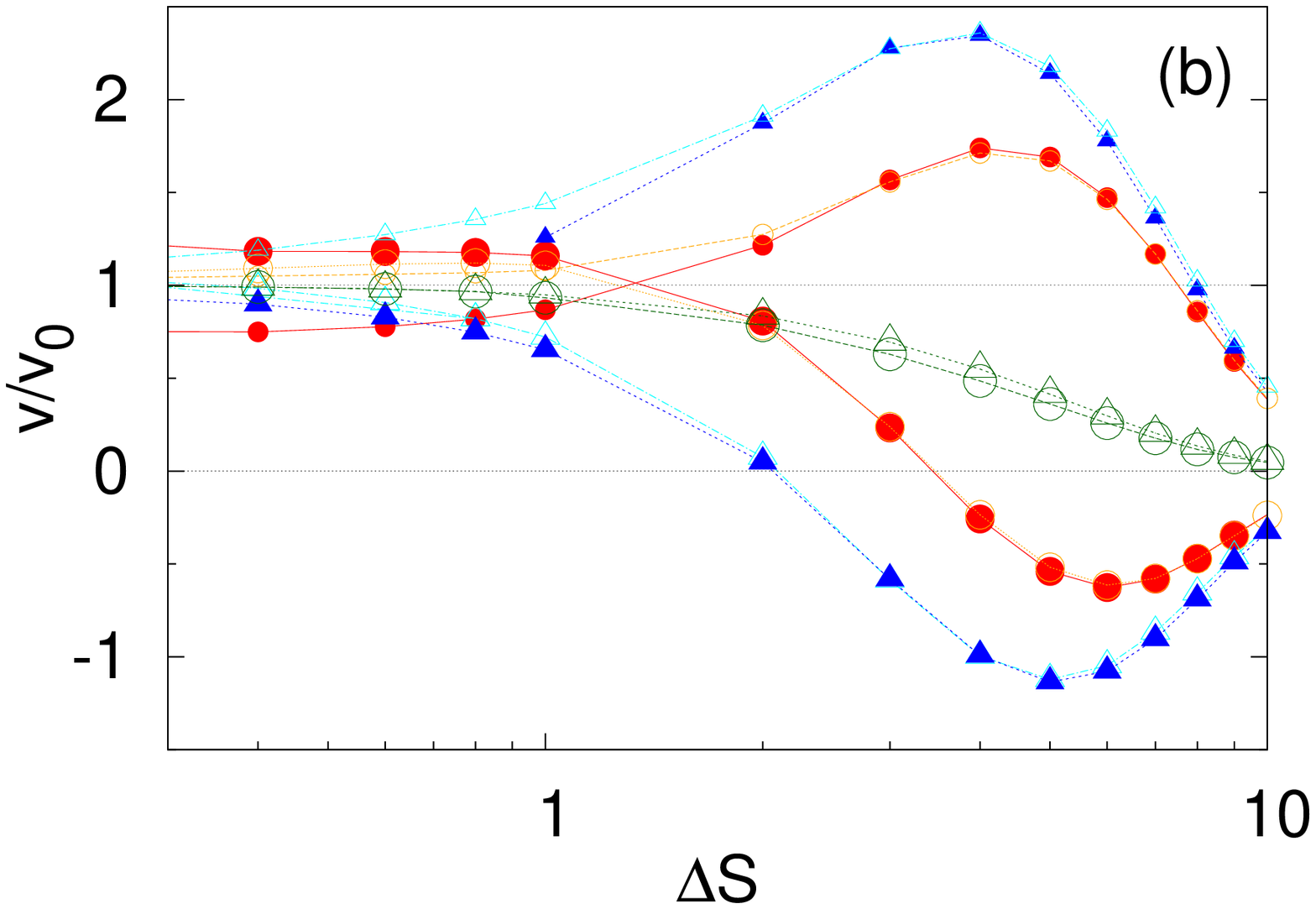}
\caption{Rectification of a processive (circles), non-processive (triangles) Brownian motor moving due to the two state model in an asymmetric channel. (a): particle velocity, in units of  $v_{0}$, for $\Delta S=0$, as a function of the phase shift $\phi_0$  for different values of the parameter $\Delta S=1.73,2.19,2.94$ (the larger the symbol   size, the larger $\Delta S$), for  $\Delta V_1=0.2$, $h_2/h_1=0.25$ and $\omega_{2,1}/\omega_{1,2}=0.01$. (b): processive (circles), non-processive (triangles) Brownian motor velocity, in units of $D_{0}/L$, as a function of $\Delta S$ and  particle radius $R$ (solid lines, with $h_0=1.25,h_1=0.2,H_2/h_1=0.25$) or $h_1$ (open points, with $R=1,h_1=1.25,h_2/h_1=0.25$) for $\phi_0=0.1,0.9$ (the larger the symbol   size, the larger $\phi_0$)  for $\mathcal{V}_1=1$ and  $\omega_{2,1}/\omega_{1,2}=0.01$. Green open circles (triangles)  represent the average velocity of processive (non-processive) motors obtained by a uniform distribution of $\phi_0$ as a function of $\
Delta S$.
}
\label{fig-twostate-asimm-asimm}
\end{figure}
\emph{Two state model.} As for the flashing ratchet case, the presence of both channel and ratchet asymmetries leads to a non trivial velocity profile upon variation of $\phi_0$. Again we find here the presence of velocity enhancement, reduction or even inversion, see fig.~\ref{fig-twostate-asimm-asimm}.a. Surprisingly, the velocity dependence on $\Delta S$ reminds the one obtained in the case of asymmetric ratchet in a symmetric channel. The entropic barrier, $\Delta S$, captures the essential response of the confined molecular motors, even better than in the cases for which the ratchet is symmetric. Finally, also the net motors flux in a disordered channel, i.e. with equally distributed $\phi_0$, has a behavior similar to the one obtained in the case of symmetric channel.

\section{VIII. Conclusions}
In this paper we have analyzed the motion of  Brownian ratchets  in confined media. We have shown that the interplay between the intrinsic ratchet motion and the geometrically-induced rectification gives rise to a variety of dynamical behaviors not observed in the absence of the geometrical confinement. A novel effect, named cooperative rectification, arises as the net result of the overlap between the dynamic induced by the ratchet and the confinement and it is responsible for the onset of net currents even when neither the ratchet nor the geometrical constraint can rectify per se. 
The dynamics of the particles can be analyzed by means of the  Fick-Jacobs  equation. Such an approach has allowed to identify   two parameters, namely the entropic barrier $\Delta S$ an the free energy drop $\Delta F$, that govern the overall dynamics of CBRs.
On one hand, $\Delta F$ controls the onset of the cooperative rectification when $\Delta F\ne 0 $. Then, cooperative rectification leads to a net current whose sign is determined by $\Delta F$. On the other hand, $\Delta S$  accounts for the relevance of  confinement  in the overall dynamics when $\Delta S \ne 0$. 

We have discussed when entropic confinement  affects  the rectification of a Brownian ratchet by contrasting physically different ratchet mechanisms. In particular, in the analysis of Brownian rectification induced by an inhomogeneous temperature profile we have clarified the relevance of the  underlying mechanism breaking detailed balance. For a flashing ratchet  the second moment of the longitudinal velocity of a CBR  differs from  the second moment associated to its transverse velocity  while  such an intrinsic anisotropy is lacking for  the thermal ratchet. As a result, confined thermal ratchets can rectify only  if there is an interplay between   the asymmetric enthalpic  potential and temperature gradients.  

Comparing the cases of a flashing ratchet and a two-state model of a molecular motor we conclude that the novel mechanism we describe is qualitatively  robust with respect to the  details of the Brownian ratchet. However, the specificity of the rectification  can affect both the  quantitative response of a Brownian ratchet to confinement and, in some cases, even affect  the qualitative  behavior observed; e.g.   velocity inversion can be observed  increasing  $\Delta S$ for  processive molecular motors in a symmetric channel while velocity inversion is never observed for non-processive motors. The Fick-Jacobs approach has provided insight to understand these qualitative differences.

We have always assumed that the confining  channel and the ratchet potential have the same period. If both components have different periodicities, or if one of them show irregularities (that can emerge, for example,  from experimental defects in the channel  build up), one can still  account for the mismatch between the ratchet potential and the channel corrugation by considering that the phase shift $\phi_0$, rather than having a well defined value, is characterized by a uniform distribution. The results reported show that, in this  situation a net current persists except for the fully symmetric geometry.

For asymmetric, intrinsically rectifying ratchets, we have seen that CBRs are very sensitive to corrugation and that  the geometrical constraints strongly affect their motion. Controlling the corrugation of the channel one can enhance significantly the net Brownian ratchet velocity or can induce velocity inversion for all the Brownian ratchet models considered. Therefore, confinement provides a means to control particle motion at small scales. Since particles with different sizes  show a differential sensitivity to the geometrical constraints, it is possible to use  channel corrugation to   segregate Brownian ratchets of different sizes, or even trap  particles. Therefore, CBRs offer new venues to particle control at small scales.

\section{Acknowledgments}
We acknowledge  the Direcci\'on General de Investigaci\'on (Spain) and DURSI project for financial support
under projects  FIS\ 2011-22603 and 2009SGR-634, respectively. J.M. Rubi acknowledges financial support from {\sl Generalitat de Catalunya } under program {\sl Icrea Academia}


\begin{thebibliography}{30}
\bibitem{footnote1}
\bibinfo{author}{A similar characterization has been made for a diffusing particle  in contact with a temperature-position heat bath in the presence of an external conservative periodic potential~\cite{Buttiker}.}
\bibitem{footnote2}
\bibinfo{author}{ In this fully symmetric regime, the velocity inversion,   upon variation of $\Delta S$, is registered only for  processive motors with $\phi_0=0.1$.}
\end{thebibliography}

\end{document}